\documentclass[aps,twocolumn,showpacs,nofootinbib,floatfix]{revtex4}

\usepackage{graphicx}

\newcommand{\cu}[1]{\chi \raisebox{-1ex}[0pt]{\scriptsize{$#1$}}}
\newcommand{\ccdot}{ \! \cdot \!}

\begin{document}

\title{Form factors  in the ``point form'' of relativistic quantum mechanics:\\
       single and two-particle currents}

\date{\today}
    
%\author{B. Desplanques\inst{1}\fnmsep\thanks{e-mail: desplanq@isn.in2p3.fr} 
%\and L. Theu{\ss}l\inst{2}\fnmsep\thanks{e-mail: Lukas.Theussl@uv.es}}
%\institute{
%Institut des Sciences Nucl\'eaires (UMR CNRS/IN2P3--UJF), 
%F-38026 Grenoble Cedex, France
%\and 
%Departamento de Fisica Teorica, Universidad de Valencia,
%E-46100 Burjassot (Valencia), Spain 
%}

\author{B. Desplanques}
\email{desplanq@lpsc.in2p3.fr}
\affiliation{Laboratoire de Physique Subatomique et de Cosmologie 
            (UMR CNRS/IN2P3--UJF--INPG), \\
             F-38026 Grenoble Cedex, France}

\author{L. Theu{\ss}l}
\email{Lukas.Theussl@uv.es}
\affiliation{Departamento de Fisica Teorica,
	     Universidad de Valencia,
	     E-46100 Burjassot, Spain}

\begin{abstract}
%\abstract{
Electromagnetic and Lorentz-scalar form factors are calculated for a 
bound system of two spin-less particles exchanging a zero-mass scalar 
particle. Different approaches are considered including solutions of a 
Bethe-Salpeter equation, a ``point form'' approach to relativistic 
quantum mechanics and a non-relativistic one. The comparison of the 
Bethe-Salpeter results, which play the role of an ``experiment'' here, with the 
ones obtained in ``point form'' in single-particle approximation, evidences 
sizable discrepancies, pointing to large contributions from two-body currents in
the latter approach. 
These ones are constructed using two constraints: ensuring current
conservation and reproducing the Born amplitude. The two-body currents so
obtained are qualitatively very different from standard ones. Quantitatively, 
they turn out not to be
sufficient to remedy all the shortcomings of the ``point form'' form factors
evidenced in impulse approximation.
%\keywords{form factors, relativity,  Wick-Cutkosky model}
%}
\end{abstract}
%
%%\PACS {{PACS-key}{discribing text of that key}}
%\PACS {
%{11.10.St} {Bound and unstable states; Bethe-Salpeter equations} --
%{13.40.Gp} {Electromagnetic form factors} --
%{12.39.Ki} {Relativistic quark model} --
%}

%\authorrunning{B. Desplanques and  L. Theu{\ss}l}
%\titlerunning{Form factors  in the ``point form'' of relativistic 
%quantum mechanics}

\maketitle

\section{Introduction}
\label{sect:1}

\noindent
The point form of relativistic quantum mechanics is much less known 
than the instant and front forms~\cite{Dirac:1949cp}, which have been 
extensively used for describing few-body systems. Recently, a calculation of 
the nucleon form factors in the former approach has revealed to be in 
surprisingly good agreement with experiment  up to 
$Q^2= 3-4 \,({\rm GeV/c})^2$~\cite{Wagenbrunn:2000es}. 
There is a slight discrepancy for the magnetic moments, 
but in view of the simplicity of the ingredients involved in the 
calculation, this is a negligible point. 

The examination  of the calculation immediately raises questions. 
It is well known phenomenology that the nucleon form factors 
at low $Q^2$ are largely dominated by the coupling of the photon 
to the nucleon via $\omega$ and $\rho$ exchanges 
(vector meson dominance~\cite{Sakurai:1960ju}). The calculation of 
ref.~\cite{Wagenbrunn:2000es} leaves no room for this important 
physical contribution, and nothing indicates that the corresponding  
phenomenology is accounted for in a hidden way by relativistic effects 
incorporated in the formalism. It is also known that effects in relation with 
the pion cloud of the nucleon explain the neutron charge 
radius~\cite{Meissner:1988ge,Acus:2000ah} 
and there is no need to invoke relativity in this respect. 
On the other hand, it is surprising that quark form factors 
(which may account for the above coupling of the $\omega$ and $\rho$ mesons) 
are discarded while 
they are needed in the construction of the quark-quark interaction 
used to describe the nucleon~\cite{Glozman:1998fs}. An attempt to incorporate 
the above physics at the quark level was recently made~\cite{Cano:2001hy}.

When looking at a given system in the Breit frame, the parameter that 
determines the form factor in ``point form'' is the velocity~\cite{Allen:2000ge}, 
$v=(Q/2M)/( 1+Q^2/4M^2)^{1/2}$, 
where $M$ is the total mass of the system. This has surprising consequences. 
It immediately follows from this expression that the charge squared radius 
scales like $1/M^2$ and therefore will increase when $M^2\rightarrow 0$. 
This is opposite to what is generally expected. A smaller mass can be obtained 
by increasing the attraction between the constituents, with the result that, 
usually, the radius decreases. 

Another way to put the problem is as follows. 
One can add an arbitrary constant to the interaction without changing the wave 
function, but changing the total mass. For the same dynamics, one would thus 
get different form factors, depending on the arbitrary constant added to the 
Hamiltonian and therefore to the mass of the system under consideration. 
This is a consequence of the formulas used in ref.~\cite{Wagenbrunn:2000es}. 
We will not discuss in detail the origin of 
this paradox and how to solve it but its very existence is a fact that cannot be 
ignored.  Notice that the problem arising in the limit $M^2\rightarrow 0$ is not 
completely academic as it applies to the pion~\cite{Amghar:2003tx}. 
A simple dimensional argument would lead to a squared radius of the order of 
$3/m_{\pi}^2 \simeq 6\,{\rm fm}^2$ (!).

Another point concerns current conservation. 
The calculation made in ref.~\cite{Wagenbrunn:2000es} does not incorporate mesonic 
exchange currents that the exchange of 
charged mesons like the pion  implies to preserve this property. Independently 
of this however, it can be checked that current conservation is not fulfilled. 
This  feature is shared by other approaches. The fact that the total momentum in 
the ``point form'' approach contains the interaction, which translates into a 
particular form of momentum conservation, could make the problem more severe 
than in the other approaches. 

The large difference between the relativistic and non-relativistic 
calculations has been attributed by the authors themselves to boost 
effects. Curiously, similar effects have not shown up in other approaches. 
To take into account the Lorentz contraction, it has been proposed 
to replace the argument of the form factor, $Q^2$, by $Q^2/(1+Q^2/4M^2)$ 
(see refs.~\cite{Friar:1973,Friar:1976kj} for a discussion). This recipe, 
only valid at small $Q^2$, gives an effect that goes in a direction 
opposite to that found in ref.~\cite{Wagenbrunn:2000es}. Calculations 
of the deuteron electro-disintegration near threshold on the light front, 
which were physically incomplete but were supposed to account for 
the various boost effects ensuring the covariance of the results, have 
evidenced no sizable effect up to 
$Q^2=10 \,({\rm GeV/c})^2$~\cite{Keister:1988,Amghar:1994te}.  
This shows that boost effects can be quite small on a large range 
of momentum transfers in some cases. It is likely that they show up only 
for some observables like the nucleon or pion form factors. In this case, 
non-relativistic estimates of the nucleon (pion) form factor are expected 
to scale like $Q^{-8}$ ($Q^{-4}$) at high momentum transfer for a 
quark-quark force which behaves like $1/r$ at small 
distances~\cite{Alabiso:1974sg}. 
The discrepancy with the QCD expectation, $Q^{-4}$ ($Q^{-2}$), 
is essentially due to a boost effect characterizing spin-1/2 
constituents, which is therefore expected 
to increase the form factor at high $Q$~\cite{Desplanques:2000ev}. 
In the ``point form'' 
calculation of the nucleon form factor performed up to now, the effect 
goes the other way round. Of course, it may show up at larger 
$Q^2$ but this does not seem to be the tendency evidenced by the results. 
A similar drop off is observed or expected in other calculations 
(see for instance ref.~\cite{Desplanques:2000ev}). The point is that these 
calculations miss  further contributions such as those of extra components 
in the light-front wave function or contact terms~\cite{Desplanques:1995ey}. 
In their absence, the nucleon and pion form factors would not have 
the correct asymptotic behavior. 

The above observations, quite puzzling in some cases, have motivated 
calculations of form factors in a simple theoretical model which 
could minimize as much as possible uncertainties due to spin 
or intrinsic form factors of the constituents~\cite{Desplanques:2001zw}. 
This model  consists of two distinguishable,
spin-less particles interacting by the exchange of a spin-less, zero-mass boson
(Wick-Cutkosky model~\cite{Wick:1954eu,Cutkosky:1954ru}). 
What accounts for the experiment 
was a calculation performed using solutions of the Bethe-Salpeter 
equation~\cite{Salpeter:1951sz}, which are easily obtained in this case. 
Form factors for the lowest bound 
states can be calculated exactly without much effort and the single-particle 
current used in the calculations ensures current conservation in all cases. 
Though it is not quite realistic, this model therefore provides 
a useful testing ground for various relativistic approaches. 
It was used  by Karmanov and Smirnov for instance to check the validity 
of the calculation of the form factors of $l=0$ and $l=1$ states 
in the light-front approach~\cite{Karmanov:1994ck}. The form factors calculated 
in this model will be referred to as ``exact'' or ``experimental'' 
ones in the following. Calculations based on the same 
``point form'' approach as mentioned above were performed using 
solutions of a mass operator reproducing the 
spectrum of the Wick-Cutkosky model. Examination of the results 
so obtained revealed that form factors were missing the ``experimental'' 
ones with two respects. The fall off of form factors is too fast 
(the power law behavior of the Born amplitude is missed) and the 
charge radius tends to be too large, especially when the binding energy 
of the system under consideration increases.

The discrepancy can be ascribed to the inadequacy of the single-particle 
current to describe the bulk of the form factors, requiring 
contributions from two-body currents which, in comparison with 
other approaches, are quite sizable~\cite{Desplanques:2001zw}. An alternative 
would consist in improving the ``point form'' implementation.   
The one mostly referred to in recent applications, also considered here, 
implies hyper-planes perpendicular to the velocity of the system. 
This feature was foreseen by Sokolov~\cite{Sokolov:1985jv}, who noticed that 
this approach is not identical to the one proposed by Dirac, which relies 
on a hyperboloid surface\footnote{The name point form was given 
by Dirac in relation with the fact that the hyperboloid surface 
is invariant under Lorentz transformations around some point, $x=0$ 
for instance. The two approaches have in common that the interaction 
is only contained in the four generators $P^{\mu}$. To emphasize the difference
with this original approach, we will use the notation with quotation marks: 
``point form''.}. 
Proposals  for improvements have been sketched in 
refs.~\cite{Desplanques:2001ze,Amghar:2002jx}.

In the present paper, we concentrate exclusively on  the first alternative. 
Results of a previous work~\cite{Desplanques:2001zw} are extended to a 
Lorentz-scalar probe 
as well as to different mass operators, to get a better 
assessment of the problems raised by the comparison of the ``point form'' 
results with the ``experimental'' ones. The sensitivity to the 
mass operator is also studied. We then consider two-body currents 
with a double aim: restore current conservation (in the case 
of an electromagnetic probe) and  reproduce the Born amplitude. 
While doing so, we faced a number of questions. Some are specific 
to the implementation of the ``point form'' approach referred to here. 
However, it turns out that other ones have a more general 
character and also occur in different forms of relativistic quantum mechanics.
Solutions that we considered could therefore be useful elsewhere after being
appropriately adapted. They have been accounted for in ref.~\cite{Amghar:2002jx}, 
which was motivated by the question of whether features evidenced by the 
``point form'' approach
were shared by the instant- and front-form ones. However, these improvements
give rise to a more complicated single-particle current, making 
the derivation of the associated two-body currents more involved. 
For the present exploratory work on these currents, we will consider 
the simplest one-body current. As will be seen, the associated  
two-body currents are already sophisticated enough.

The plan of the paper is as follows. The second section is devoted 
to reminding the impulse-approximation expressions of the form factors 
we are calculating and also includes the  case of a scalar probe. 
In the third section, we extend numerical results presented 
in a previous paper~\cite{Desplanques:2001zw} to the scalar form factor and to
other mass operators. The fourth section is concerned with general comments 
inspired by the results so obtained. It deals with both the low momentum 
transfers, where current conservation is an important constraint, and 
high momentum ones, where the consideration of the Born amplitude provides an 
important benchmark. Results involving two-body currents are given and 
discussed in the fifth section. Many expressions pertinent to the present work 
are gathered in the appendix.    
 
%%%%%%%%%%%%%%%%%%%%%%%%%%%%%%%%222222222222222%%%%%%%%%%%%%%%%%%%%%%%%%%%%%%%%

\section{Form factors in different formalisms}
\label{sect:2}

\noindent
Extending our previous work~\cite{Desplanques:2001zw}, we here consider form 
factors relative to both a scalar 
and an electromagnetic probe. Quite generally, the corresponding matrix element 
between two states with $l=0$, possibly different, may be expressed as:
\begin{eqnarray}
\sqrt{2E_f2E_i} \, \left<f|J^{\mu}|i\right> &=& F_1(q^2)
(P^{\mu}_f+P^{\mu}_i) + F_2(q^2)q^{\mu},
\nonumber \\
\sqrt{2E_f2E_i} \, \left<f|S|i\right>&=& \, F_0(q^2) \, (4m),
\label{2a}
\end{eqnarray}
where $q^{\mu}=P^{\mu}_f-P^{\mu}_i$ and $q^2=-Q^2$. The operators $S$ and 
$J^{\mu}$ on the 
l.h.s. of eq.~(\ref{2a}), describe the interaction with the external probe, 
respectively of Lorentz-scalar and vector types. Due to current conservation,  
the form factors $F_1(q^2)$ and $F_2(q^2)$ have to fulfill the following 
relationship:
\begin{equation}
F_1(q^2)\,(M_f^2-M_i^2) + F_2(q^2)\,q^2= 0.
\label{2b}
\end{equation}
For an elastic process,  this is automatically fulfilled since 
$F_2(q^2)$ vanishes identically from symmetry arguments 
alone, but this does not imply that current conservation holds at 
the operator level, as it should. For an inelastic process, eq.~(\ref{2b})
implies that $F_1(q^2)\rightarrow 0$ with $q^2$.

The normalization of the form factors in eq.~(\ref{2a}) is for some 
part arbitrary. Assuming that the system under consideration is made of one 
charged and one neutral particle, it is appropriate to normalize $F_1(q^2)$ such 
that $F_1(q^2=0)=1$. In absence of a conservation law for the scalar probe, we 
normalize the scalar form factor such that $F_0(q^2)$ and $F_1(q^2)$ coincide in 
the non-relativistic limit. Even so, other normalization factors with the same 
non-relativistic limit in eq.~(\ref{2a}) could be chosen, such as $2M$ instead 
of $4m$, but we did not find any compelling reason to do it 
(see discussion in sect.~\ref{sect:3}). 

\begin{figure}[tb]
\begin{center}
\includegraphics[width=20em]{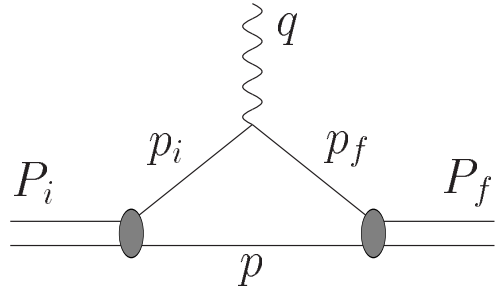}
\end{center}
\caption{Representation of a scalar particle or virtual photon absorption on a 
two-body system, with indication of the kinematical definitions.\label{fig1}}
\end{figure}

In the following, we successively consider form factors in the Wick-Cutkosky 
model and in the ``point form'' approach. Some of the matter given in an earlier 
paper~\cite{Desplanques:2001zw} is provided in appendices~\ref{app:a} 
and~\ref{app:c} together with a few formulas. The 
contribution  we intend to calculate is shown in Fig.~\ref{fig1}. 

%%%%%%%%%%%%%%%%%%%%%%%%%%%%%%%2.1
\subsection{Form factors in the Wick-Cutkosky model}
\label{subsec:21}

\subsubsection{Interaction and Bethe-Salpeter amplitudes}

\noindent
What will account for our ``experiment'' is based on the Wick-Cutkosky model. 
The Bethe-Salpeter amplitudes take in this case a relatively 
simple integral form for the lowest state with a given angular momentum $l$:
\begin{equation}
 \cu{P}( p ) =\int dz\,\frac{g_n(z) \, {\mathcal Y}_l^m(\vec{p}\,)}{(m^2 - 
\frac{1}{4}P^2- p^2-z\,P\ccdot p-i\epsilon)^{n+2}},
\label{2d}
\end{equation}
with $n=l+1$.
For the first radial excitation, the Bethe-Salpeter amplitude reads
\begin{equation}
 \cu{P}( p ) =\int dz\,\frac{g_n(z) \, 
  (p^2+m^2-\frac{1}{4}P^2)\, {\mathcal Y}_l^m(\vec{p}\,)}{
  (m^2 - \frac{1}{4}P^2- p^2-z\,P\ccdot p-i\epsilon)^{n+2}},
\label{2e}
\end{equation}
where now $n=l+2$.
In these expressions, ${\mathcal Y}_l^m(\vec{p}\,)=|\vec{p}\,|^lY_l^m(\hat{p})$ 
and $g_n(z)$ is a solution of the second order differential 
equation~\cite{Wick:1954eu,Cutkosky:1954ru}:
\begin{eqnarray}
\label{2f}
\nonumber
\lefteqn{
(1-z^2)\,g_n''(z)  +  2(n-1) \, z \, g_n'(z) - n(n-1)\,g_n(z)}\\
&& \qquad \qquad \qquad \, + 
\frac{\alpha}{\pi (\epsilon^2 \, z^2 + 1 - \epsilon^2)}\,g_n(z)=0,
\end{eqnarray}
with $\epsilon^2=P^2/(4m^2)$ and the boundary conditions 
$g_n(z= \pm 1)=0$. Only normal solutions (without node) are considered here. 
In the small binding limit, the function $g_1(z)$ of the ground state 
is given by $1-|z|$, while in the deep binding limit
($M=0$), $g_1(z)\propto1-z^2$. 

\subsubsection{Expressions of form factors}

\noindent
For the model under consideration here, the general (and exact) expression of 
the matrix element of the current can be written in terms of the Bethe-Salpeter 
amplitudes
\begin{eqnarray}
\label{2c}
\nonumber
\lefteqn{
\sqrt{2E_f\,2E_i} \,\left<f|S|i\right> = i\int \frac{d^4p}{(2\pi)^4} \, (2m)} \\
\nonumber &&
\qquad \quad \cu{P_f}(\frac{1}{2} P_f-p ) \, (p^2-m^2)\, 
\cu{P_i}( \frac{1}{2} P_i-p ), \\
\nonumber &&
\lefteqn{
\sqrt{2E_f\,2E_i}\,\left<f|J^{\mu}|i\right> =i\int \frac{d^4p}{(2\pi)^4} 
\left( P^{\mu}_f+P^{\mu}_i-2\,p^{\mu}\right) }   \\
&&
\qquad \quad \cu{P_f}(\frac{1}{2} P_f-p ) \,
\,(p^2-m^2)\,\, \cu{P_i}( \frac{1}{2} P_i-p ).\qquad
\end{eqnarray}
According to our normalization convention, a factor $2\,m$ has been 
introduced in the above expression of the scalar form factor. In a scalar 
theory, like the Wick-Cutkosky model, this factor is often separated out to make 
the coupling constant dimensionless and directly comparable to $\alpha_{QED}$. 

Using eqs.~(\ref{2d}) and~(\ref{2e}), the calculation of the matrix element, 
eq.~(\ref{2a}), can be partly performed by employing the Feynman 
method. A few expressions of interest here are given 
in appendix~\ref{app:a} for both elastic and inelastic cases. 
Though the calculation is not straightforward, it can be checked 
that the current conservation, eq.~(\ref{2b}), is verified 
in the inelastic case.

%%%%%%%%%%%%%%%%%%%%%%%%%%2.2
\subsection{Form factors in the ``point form'' approach}
\label{subsec:22}

\noindent
The ``point form'' approach to relativistic quantum mechanics is characterized 
by the property that, among the 10 generators of the Poincar\'e group, 
only the four momenta $P^{\mu}$ contain the interaction. These ones 
can be written as the sum of the free particle and interaction contributions:
\begin{equation}
P^{\mu}=P^{\mu}_{free}+P^{\mu}_{int}=M\, V^{\mu}.
\label{2g}
\end{equation}
where the last equality defines the four-velocity operator $V^{\mu}$ in terms of
the mass operator $M=\sqrt{P^2}$. 

In the implementation of the ``point form'' employed in recent applications, 
the simplest choice compatible with the Poincar\'e algebra 
has been made for the interaction part of the four-momentum. 
This one assumes the form:
\begin{equation}
P^{\mu}_{int}=M_{int}\, V^{\mu},
\label{2h}
\end{equation}
which, together with eq.~(\ref{2g}), implies:
\begin{equation}
 V^{\mu} \,(M-M_{int})=V_{free}^{\mu} \,M_{free},
\label{2i}
\end{equation}
where $M_{free}=\sqrt{P^2_{free}}$. From this one obtains:
\begin{equation}
\label{2k}
V^{\mu} = V_{free}^{\mu},\label{2j} \qquad M = M_{free}+M_{int}.
\end{equation}
The last relation, which could also be used as a definition of $M_{int}$, is 
consistent with the choice of eq.~(\ref{2h}).

The form of the interaction part  
of the four-momentum considered in eq.~(\ref{2h}) can be associated 
to a physics description on a hyper-plane perpendicular 
to the four-velocity of the system, an observation made previously 
by Sokolov~\cite{Sokolov:1985jv}. As for the interaction term $M_{int}$, 
it may be chosen according to some theoretical prejudice or to reproduce 
some experimental spectrum as done in ref.~\cite{Wagenbrunn:2000es}. 
Instead of $M$, one can also use the square of the operator~\cite{Allen:2000ge}. 
This has some advantage since, in the two-body case, $M^2$ is very close to a 
Schr\"odinger equation. The solutions of this one may therefore also be used 
as was done in ref.~\cite{Desplanques:2001zw}. In this case, it turns out that 
the theoretical spectrum obtained 
with a Coulomb-like potential reproduces rather well the one of the 
Wick-Cutkosky model. We have thus a set of analytic wave functions that can be 
used for the calculation of form factors. However, while doing this, one has to 
worry that the currents associated to the mass operators $M^2$ and $M$ may not 
be the same (a conserved current is more easily built in one case than in the 
other). 

\subsubsection{Mass operator and solutions}

\noindent
In this work, we will still refer to the above calculations made with the 
Coulomb-like potential (denoted model v0, see app.~\ref{app:c}) 
but we will also consider 
solutions of a linear mass operator. Consistently with eq.~(\ref{2k}), the
corresponding equation takes the form:
\begin{equation}
M \, \psi(k)=2 \, e_k \, \psi(k) + \int  \frac{d \vec{k}\,'}{(2\pi)^3} \, 
V_{int}(\vec{k},\vec{k}\,')\,  \psi(k'),
\label{2m}
\end{equation}
where $e_k=\sqrt{m^2+k^2}$, while $\vec{k}$ represents an internal 
variable that, in the non-relativistic limit, could be identified 
with the relative momentum.  Without certitude about 
which approach is the best, this will give insight on the related 
uncertainty. The main difficulty is to derive an interaction 
to be used in eq.~(\ref{2m}) such that it reproduces the spectrum 
of the Wick-Cutkosky model. The first order term one can think 
of is motivated by a standard field-theory approach to the derivation 
of a one-boson interaction. It is given by:
\begin{equation}
\label{2mp}
V_{int}(\vec{k},\vec{k}\,')= -\frac{m}{e_k}\, \frac{g^2}{ 
(\vec{k}-\vec{k}\,')^2}\,\frac{m}{e_{k'}},\label{2n}
\end{equation}
This model, denoted v1, 
which is non local, misses however properties of the Wick-Cutkosky model 
that were reproduced by the  Coulomb-like potential, such as the degeneracy 
of $1p$ and $2s$ states. Relying on the fact that the square 
of the mass operator should be close to the  one which works 
($(2e_k+M_{int})^2= 4e_k^2-4m\frac{\alpha_{eff}}{r}$ in configuration 
space), an extra term can be derived:
\begin{eqnarray}
\label{2o}
\nonumber
\lefteqn{
\Delta V_{int}(\vec{k},\vec{k}\,')= -\frac{m}{e_k}
\left( \frac{g^2}{ (\vec{k}-\vec{k}\,')^2}\, 
\frac{2\,e_k\,e_{k'}-m\,(e_k+e_{k'})}{m\,(e_k+e_{k'})} 
\right. } \\&& \nonumber
+ \left. \frac{g^4}{32\,m\,|k-k'|} \, \frac{8\,m^3}{(e_k+e_{k'})(e_k+m) 
(e_{k'}+m)} \right)  \frac{m}{e_{k'}}. \\
\end{eqnarray}
The expression is exact for the part linear in $g^2$ and includes 
corrections at the order $g^4$ in an approximate way (exact 
in the lowest $1/m$ order with correct asymptotic $1/k^4$ power 
law (up to log terms)). The addition of the above correction 
to the interaction given in eq.~(\ref{2n}) defines an improved 
model, denoted v2.

\subsubsection{Expressions of form factors}

\noindent
Solutions of the mass operator $M$ can now be used for the calculation 
of form factors. This was described in ref.~\cite{Allen:2000ge} for the matrix 
element of the single-particle current. However, instead of using expressions 
where appropriate boosts have to be performed, we rely on expressions 
whose Lorentz-covariance is explicit:
\begin{eqnarray}
\label{2p} 
\nonumber
\lefteqn{ 
\sqrt{2E_f\,2E_i} \,\left<f|S|i\right> = 
\sqrt{2M_f\,2M_i} \, \frac{1}{(2\pi)^3 }} \\ && \nonumber  
\int d^4p \, d^4p_f \,  d^4p_i \, d\eta_f \, d\eta_i \,
\sqrt{(p_f+p)^2 \, (p_i+p)^2 }\\ && \nonumber
\delta(p^2-m^2) \,  \delta(p_f^2-m^2) \, \delta(p_i^2-m^2) \,
\theta( \lambda_f \ccdot   p_f) \, \theta(\lambda_f \ccdot   p) \\ && \nonumber
\times \, \theta(\lambda_i \ccdot   p)\, \theta(\lambda_i \ccdot   p_i)
\,\delta^4(p_f+p-\lambda_f \eta_f) \,
\,\delta^4(p_i+p-\lambda_i \eta_i)   \\ && 
\qquad\times \,\phi_f((\frac{p_f-p}{2})^2)\, \phi_i((\frac{p_i-p}{2})^2) \, 
(2m), 
\end{eqnarray}
\begin{eqnarray}
\label{2q} 
\nonumber
\lefteqn{ 
\sqrt{2E_f\,2E_i} \,\left<f|J^{\mu}|i\right> = 
\sqrt{2M_f\,2M_i} \, \frac{1}{(2\pi)^3 }} \\ && \nonumber
\int d^4p \, d^4p_f \,  d^4p_i \, d\eta_f \, d\eta_i \,
\sqrt{(p_f+p)^2 \, (p_i+p)^2 }
 \\ && \nonumber
\delta(p^2-m^2) \,  \delta(p_f^2-m^2) \, \delta(p_i^2-m^2) \,
\theta( \lambda_f \ccdot   p_f) \, \theta(\lambda_f \ccdot   p) \\ && \nonumber
\times \, \theta(\lambda_i \ccdot   p)\, \theta(\lambda_i \ccdot   p_i)
\delta^4(p_f+p-\lambda_f \eta_f) \,
\delta^4(p_i+p-\lambda_i \eta_i)   \\ && 
\qquad\times \,\phi_f((\frac{p_f-p}{2})^2)\, \phi_i((\frac{p_i-p}{2})^2) \,
 (p_f^{\mu}+ p_i^{\mu}),
\end{eqnarray}
where $\lambda^{\mu}_{i,f}$ are unit four-vectors
proportional to the four-momenta of the total system in the initial and 
final states, $\lambda^{\mu}_i = P^{\mu}_i/M_i$ and
$\lambda^{\mu}_f= P^{\mu}_f/M_f$. These four-vectors can be expressed in 
terms of the corresponding velocities, $\lambda^0=(\sqrt{1-v^2})^{\,-1}$ and 
$\vec{\lambda}= \vec{v}\,(\sqrt{1-v^2})^{\,-1}$.

Except obviously for the current that behaves like a four-vector,  
all quantities in the above expressions are Lorentz-invariant. 
This is achieved by the introduction of auxiliary variables $\eta_i$ 
and  $\eta_f$, which play the role of an off-energy shell invariant mass. 
When they are integrated over, they give rise 
to the following three-dimensional $\delta$ functions:
\begin{equation} 
\delta
\Big(\vec{p}_i+\vec{p}-\frac{\vec{\lambda}_i}{\lambda_i^0}(p^0_i+p^0)\Big) 
\,\, {\rm and} \,\, 
\delta
\Big(\vec{p}_f+\vec{p}-\frac{\vec{\lambda}_f}{\lambda_f^0}(p^0_f+p^0)\Big).
\label{2r}
\end{equation}
These relations are pertinent to the ``point form'' approach referred to 
throughout this paper  and account for the fact that the velocity $\vec{v}$, 
defined as the ratio of the sum of the momenta $\sum \vec{p}_j$ 
and the sum of the kinetic energies $\sum e_j$, is conserved 
for a given system~\cite{Allen:2000ge}. They replace the conservation of 
momenta in the instant-form approach. It has not been possible 
to show that they strictly follow from describing physics 
on a hyperboloid surface~\cite{Klink:2000pp}. Instead, they can be obtained 
when this surface is taken as a hyper-plane orthogonal to the four-velocity 
of the system, consistently with the form of eq.~(\ref{2h}) and the 
observation made by Sokolov~\cite{Sokolov:1985jv}. On the other hand, it can be 
checked that, in the c.m., the wave function $\phi$ only depends 
on the relative momentum of the two particles. Moreover, by direct 
integration or after performing a change of variable, one recovers 
that the current of a given system is given by 
$( \langle J^0\rangle, \vec{\langle J\rangle})  = (1, \vec{v})$, 
in agreement with the standard normalization of the wave function, 
$ \int\frac{d\vec{k}}{(2\pi)^3} \,\phi^2(\vec{k})=1$. We will come back 
to this normalization in sect.~\ref{sect:5}, when considering two-body currents. 

The form factors we are interested in, $F_0(q^2)$, $F_1(q^2)$ and $F_2(q^2)$, 
can be calculated from eqs.~(\ref{2p},\ref{2q}) in any frame. However, they 
take a simpler expression in the Breit frame, defined by  
$\vec{v}=\vec{v}_f=-\vec{v}_i$, with $v$ expressed in terms of the  momentum 
transfer $Q$: $v^2=\frac{Q^2+(M_f-M_i)^2}{Q^2+(M_f+M_i)^2}$.  The 
electromagnetic form factors are more appropriately expressed in terms of 
auxiliary quantities, $\tilde{F}_1(q^2)$ and  $\tilde{F}_2(q^2)$, which 
involve the time and spatial parts of the current, respectively. We thus have: 
\begin{eqnarray}
\label{2s}
F_0(q^2)& =&  \frac{\sqrt{M_f\,M_i}}{2\,m} \int \frac{d \vec{p}}{(2\pi)^3} 
\,\phi_f(\vec{p}_{tf}) \, \frac{m}{e_p} \, \phi_i(\vec{p}_{ti}), \nonumber \\
\tilde{F}_1(q^2)& = & \frac{1+v^2}{\sqrt{1-v^2}} \int \frac{d \vec{p}}{(2\pi)^3} 
\,\phi_f(\vec{p}_{tf}) \,\phi_i(\vec{p}_{ti}), \nonumber \\
\tilde{F}_2(q^2) \, \vec{v}& =&  -\frac{1+v^2}{\sqrt{1-v^2}}    
\int \frac{d \vec{p}}{(2\pi)^3} \phi_f(\vec{p}_{tf})   \frac{\vec{p}}{e_p} 
\phi_i(\vec{p}_{ti}),
\end{eqnarray}
together with  
\begin{eqnarray}
\label{2t}
\nonumber 
\lefteqn{
F_1(q^2) \,  \sqrt{2M_f\,2M_i}   =} \\ \nonumber && \quad
  \tilde{F}_1(q^2) \, (M_f+M_i) - \tilde{F}_2(q^2) \, (M_f-M_i)  \\ \nonumber
  \lefteqn{
F_2(q^2) \,  \sqrt{2M_f\,2M_i}  =} \\  &&   \quad
-\tilde{F}_1(q^2) \, (M_f-M_i) + \tilde{F}_2(q^2)\, (M_f+M_i).
\end{eqnarray}
The (Lorentz-) transformed momenta are defined as:
$ (p^x,p^y,p^z)_{ti,f}= (p^x,p^y, \frac{p^z \pm v \, e_p}{\sqrt{1-v^2}})$, 
together with $e_p=\sqrt{m^2+\vec{p}^{\,2}}$. 

Contrary to eq.~(\ref{2c}), there is no guarantee that current 
conservation, eq.~(\ref{2b}), is fulfilled by eqs.~(\ref{2s}, 
\ref{2t}). How much it is violated for inelastic transition is of interest. 

%%%%%%%%%%%%%%%%%%%%%%%%%%2.3
\subsection{Non-relativistic form factors}
\label{subsec:23}

Finally, we recall the non-relativistic expressions of the elastic 
and inelastic form factors $F_0(q^2)$, $F_1(q^2)$ and $F_2(q^2)$, 
that can be calculated with the same wave functions as used 
in eqs.~(\ref{2s}). For a local interaction model, like v0, where 
the simplest single-particle current is conserved, they read:
\begin{eqnarray}
\label{2u}
\nonumber 
F_0(q^2) =  F_1(q^2)= 
\int \frac{d \vec{p}}{(2\pi)^3} \,\, 
\phi_f(\vec{p}-\frac{1}{4}\vec{q})
\,\, \phi_i(\vec{p}+\frac{1}{4}\vec{q}),  \\
F_2(q^2) \, \frac{\vec{q}}{4m} = -\int \frac{d \vec{p}}{(2\pi)^3} 
\phi_f(\vec{p}-\frac{\vec{q}}{4}) \frac{\vec{p}}{m} 
\phi_i(\vec{p}+\frac{\vec{q}}{4}).\qquad
\end{eqnarray}
In this case, it can be checked that the form factors verify the 
current-conservation condition, eq.~(\ref{2b}). 

In a few cases, 
form factors involving Coulombian wave functions can be calculated 
analytically. Their expression is given in appendix~\ref{app:c}. 
For a non-local interaction model like v1, which  includes a 
semi-relativistic kinetic energy or normalization factors $m/e$, 
the expression of the form factors may involve slightly different 
single-particle operators while preserving the Galilean invariance. 
These ones, which are model dependent, will be given later on 
together with  the two-body currents that are then necessary 
to fulfill current conservation. These
form factors will serve as a useful benchmark for comparison 
with the ``point form'' results, which should represent an improvement with
respect to the ``exact'' ones.

%%%%%%%%%%%%%%%%%%%%%%%%%%%%%%%33333333333333%%%%%%%%%%%%%%%%%%%%%%%%%%%%%%%%%%

\section{Results in impulse approximation} 
\label{sect:3}

\noindent
In this section, we complete results obtained  in an earlier 
paper for electromagnetic form factors~\cite{Desplanques:2001zw}
by providing scalar ones. They should allow one to get a better insight 
on how the ``point form'' approach does with respect to the 
other ones. Results corresponding to different mass operators 
are also presented.

\begin{table}[tb]
\caption{Elastic form factor $F_1(q^2)$ for the ground state: 
Non-relativistic ($N.R.$) and ``point form'' ($P.F.$) calculations 
are performed with Coulombian wave functions 
(model v0, see app.~\protect\ref{app:c}). 
The binding energies of the states, in units of the constituent
mass m, are given by: 
$E=0.0842$ ($\alpha=1$), $E=0.432$ ($\alpha=3$) and $E=2.0$ ($\alpha=2\pi$). 
In the last case, results for two slightly different values of $E$ 
are given for the ``point form'' results (see explanation in the text).}
\label{t10} 
\begin{ruledtabular}
\begin{tabular}{lccccc}
  $Q^2/m^2$             &   0.01   &   0.1  &  1.0   &  10.0 & 100.0
  \\ [1.ex] \hline
 $\alpha=1$             &         &        &        &         &  \\ [0.ex] 
 $B.S.$     & 0.984    & 0.856  & 0.309  &  0.137-01  & 0.213-03  \\ [0.ex] 
 $N.R.$     & 0.985    & 0.864  & 0.323  &  0.136-01  & 0.169-03  \\ [0.ex]
 $P.F.$     & 0.984    & 0.853  & 0.299  &  0.974-02  & 0.343-04  \\  [1.ex]    

 $\alpha=3$            &          &        &        &         &  \\ [0.ex] 
 $B.S.$     & 0.996    & 0.962  & 0.705  &  0.139  & 0.503-02  \\ [0.ex] 
 $N.R.$     & 0.997    & 0.968  & 0.740  &  0.146  & 0.338-02  \\ [0.ex]
 $P.F.$     & 0.995    & 0.949  & 0.621  &  0.563-01  & 0.228-03  \\  [1.ex]    

 $\alpha=2\pi$         &          &           &          &         &  \\ [0.ex] 
 $B.S.$        & 0.998  & 0.983     & 0.848    &  0.339  & 0.285-01   \\ [0.ex] 
 $N.R.$        & 0.999  & 0.988     & 0.886    &  0.379  & 0.190-01   \\ [0.ex]
 $P.F._{(E=1.90)}$& 0.614  & 0.398-01  & 0.111-03 &  0.126-06 & 0.127-09 \\ [0.ex]    
 $P.F._{(E=1.95)}$& 0.187  & 0.143-02  & 0.193-05 &  0.199-08 & 0.200-11 \\ [1.ex]  
\end{tabular}
\end{ruledtabular}
\end{table}

Results are presented successively in three tables: for the elastic 
form factor, $F_1(q^2)$ (table~\ref{t10}), for the elastic form 
factor, $F_0(q^2)$ (table~\ref{t20}), and for an inelastic transition 
from the ground to the first radially excited state (table~\ref{t30}). 
In the two first cases, three values of the coupling constant have been 
considered: $\alpha=1$, $\alpha=3$ and $\alpha=2\pi$, where  $\alpha$ is related
to the coupling constant $g^2$ used in eq.~(\ref{2mp}) by $\alpha=g^2/(4\,\pi)$. 
These values correspond to a small, a moderate  and a large
binding energy (4\%, 20\% and 100\% of the total mass of the constituents),
respectively. 
The last value is an extreme one since the total mass is zero but, 
as it sometimes happens, such cases better reveal features pertinent 
to some approach. In the zero-mass case, results for the ``point form'' approach 
essentially vanish at $Q^2 \neq 0$ ($F_1$) or even identically ($F_0$). 
For this reason, the corresponding results are given for two values 
of the binding energy, $E=1.90\,m$ and $E=1.95\,m$, which allow one 
to approach the limit $M=0$, while at the same time results obtained 
with the Bethe-Salpeter or non-relativistic approaches are 
essentially unchanged. For the inelastic transition, results 
are presented for the three form factors $F_0(q^2)$, $F_1(q^2)$ 
and $F_2(q^2)$, and for one value of the coupling constant, 
$\alpha=3$. 

\subsection{Elastic charge form factors}
\label{subsec:31}

\noindent
Results for the elastic form factor, $F_1(q^2)$ (table~\ref{t10}), 
have been already discussed in ref.~\cite{Desplanques:2001zw}. 
As noticed there,  the non-relativistic calculation agrees 
relatively well with the ``exact'' results. 
Reproducing at the same time the low momentum range, constrained 
by the charge associated to the conserved current, and the high 
momentum range, constrained by the Born amplitude, the form factor 
$F_1(q^2)$, calculated in the non-relativistic approach, cannot 
be wrong by a large amount (up to log terms). 
As noticed in ref.~\cite{Desplanques:2001zw},  
the ``point form'' approach departs from the ``exact'' result 
at the highest values of $Q^2$ that were considered. Results at  
$Q^2=100\,m^2$ clearly show the failure of this approach in the impulse
approximation  for a large coupling ($\alpha=2\pi$),  but also at 
small couplings. This is obviously due to the asymptotic behavior 
of the form factor that varies like $1/Q^6$ instead of $1/Q^4$ for the 
``exact'' and non-relativistic calculations. As for the tendency 
of the ``exact'' results to depart from the non-relativistic ones 
at these high $Q^2$ values, it signs the onset of log term corrections 
in the former ones.

\subsection{Elastic scalar form factors}
\label{subsec:32}

\noindent
Results for the scalar form factor $F_0(q^2)$ (table~\ref{t20})
confirm the overall agreement of the non-relativistic calculation 
with the ``exact'' one. Contrary to $F_1(q^2)$, some discrepancy 
appears at $Q^2=0$. In absence of a conserved charge in this case, 
this points to the role of relativistic corrections 
at low $Q^2$. These ones remain moderate however, including 
the extreme case $\alpha=2\pi$ ($M=0$). In comparison, for the 
``point form'' results the discrepancy 
with the ``exact'' ones is larger than for $F_1(q^2)$,
both at small and large $Q^2$. At low $Q^2$, part of the effect 
is due to the factor $M/(2m)$ appearing in eq.~(\ref{2s}), 
which has no counterpart in the other approaches. This effect becomes 
especially large when approaching the limit $M=0$, a result 
which is independent of the way the scalar form factor is defined in eq. 
(\ref{2a}). The ratio of the $F_0(q^2)$ and $F_1(q^2)$ form factors 
however depends on this definition, requiring some caution about the 
conclusion that their comparison can suggest. As there is a close 
relationship between these form factors in the Wick-Cutkosky model, 
both numerically and algebraically\footnote{The two form factors
$F_0(q^2)$ and $F_1(q^2)$ are equal in the Born approximation 
and from higher orders, one expects log corrections leading 
to $F_1(q^2)=2\,F_0(q^2)$ in the ultra-relativistic 
domain.}, we are rather tempted to think that, in the ``point form'' 
approach,  the form factor $F_0(q^2)$ is strongly suppressed 
with respect to $F_1(q^2)$ at low $Q^2$ in the limit $M \rightarrow 0$. 
At high $Q^2$, the form factor rather scales like $1/Q^8$, 
instead of $1/Q^4$, hence a larger discrepancy with the other 
results than for $F_1(q^2)$. 

\begin{table}[tb]
\caption{Elastic form factor $F_0(q^2)$ for the ground state: Same as in 
table~\protect\ref{t10}.}
\label{t20}
\begin{ruledtabular}
\begin{tabular}{lccccc}
  $Q^2/m^2$             &   0.01   &   0.1  &  1.0   &  10.0 & 100.0
  \\ [1.ex] \hline
 $\alpha=1$             &         &        &        &         &  \\ [0.ex] 
 $B.S.$     & 1.024    & 0.887  & 0.313  &  0.128-01  & 0.18-03  \\ [0.ex] 
 $N.R.$     & 0.985    & 0.864  & 0.323  &  0.135-01  & 0.17-03  \\ [0.ex]
 $P.F.$     & 0.949    & 0.813  & 0.256  &  0.43-02  & 0.33-05  \\  [1.ex]    

 $\alpha=3$            &          &        &        &         &  \\ [0.ex] 
 $B.S.$     & 1.123    & 1.080  & 0.767  &  0.132  & 0.394-02  \\ [0.ex] 
 $N.R.$     & 0.996    & 0.968  & 0.740  &  0.145  & 0.337-02  \\ [0.ex]
 $P.F.$     & 0.682    & 0.641  & 0.366  &  0.16-01  & 0.15-04  \\  [1.ex]    

 $\alpha=2\pi$         &          &           &          &         &  \\ [0.ex] 
 $B.S.$        & 1.247  & 1.222     & 1.016    &  0.338  & 0.217-01   \\ [0.ex] 
 $N.R.$        & 0.997  & 0.987     & 0.885    &  0.378  & 0.189-01   \\ [0.ex]
 $P.F._{(E=1.90)}$& 0.175-01  & 0.435-03  & 0.28-06 &  0.54-10 & 0.79-14 \\ [0.ex]    
 $P.F._{(E=1.95)}$& 0.162-02  & 0.335-05  & 0.86-09 &  0.14-12 & 0.18-16 \\ 
[1.ex]    
\end{tabular}
\end{ruledtabular}
\end{table}

\begin{table}[tb]
\caption{Inelastic form factors $F_0(q^2)$, $F_1(q^2)$ and $F_2(q^2)$, 
for a transition from the ground state to the first radially excited one:
Non-relativistic and ``point form'' results are obtained with Coulombian wave
functions (model v0), as in tables~\protect\ref{t10} and~\protect\ref{t20}. 
The results correspond to  $\alpha=3$ ($E_i=0.4322\,m$, $E_f=0.1036\,m$ 
for $B.S.$ and $0.098\,m$ for $N.R.$ and $P.F.$)}
\label{t30}
\begin{ruledtabular}
\begin{tabular}{lccccc}
  $Q^2/m^2$     &   0.01   &   0.1  &  1.0   &  10.0 & 100.0 \\ [1.ex] \hline
 $B.S.$   &   &   &        &       &  \\ [0.ex] 
 $F_0$  & 0.538-01 & 0.781-01  & 0.172-00  & 0.537-01 & 0.163-02\\ [0.ex] 
 $F_1$  & 0.032-01 & 0.298-01  & 0.145-00  & 0.584-01 & 0.214-02\\ [0.ex] 
 $F_2$  & 0.369-00 & 0.340-00  & 0.165-00  & 0.665-02 & 0.217-04\\ [1.ex]

 $N.R.$  &   &   &        &      &   \\ [0.ex] 
 $F_0$  & 0.032-01 & 0.296-01  & 0.151-00  &  0.550-01 & 0.121-02\\ [0.ex] 
 $F_1$  & 0.032-01 & 0.296-01  & 0.151-00  &  0.550-01 & 0.121-02\\ [0.ex] 
 $F_2$  & 0.369-00 & 0.342-00  & 0.174-00  &  0.636-02 & 0.140-04\\ [1.ex]

 $P.F.$   &   &   &        &       &  \\ [0.ex] 
 $F_0$  & - 0.046-01 &  0.171-01 & 0.090-00  & 0.099-01  &  0.119-04\\ [0.ex] 
 $F_1$  &   0.101-01 &  0.372-01 & 0.140-00  & 0.283-01  &  0.139-03\\ [0.ex] 
 $F_2$  &   0.324-00 &  0.293-00 & 0.119-00  & 0.217-03  & -0.118-04\\ [1.ex]
\end{tabular}
\end{ruledtabular}
\end{table}

\subsection{Inelastic form factors}
\label{subsec:33}

\noindent
When making a comparison with the ``exact'' results, 
it was noticed in ref.~\cite{Desplanques:2001zw} that, for the ``point form'' results, 
a relative change in sign of the form factors $F_1(q^2)$ and $F_2(q^2)$ 
occurs, preventing from fulfilling the current conservation 
constraint given by eq.~(\ref{2b}). The fact that the non-relativistic 
calculation does better than the ``point form'' one is confirmed 
by results for the scalar form factor, $F_0(q^2)$. In particular, at 
low $Q^2$ the first one has the right sign while the other one has not. 
However, the discrepancy in size is large in both cases. Much better 
than the elastic case, the present results evidence the role 
of relativistic corrections. Anticipating on the next section, 
let us mention that a non-zero value of $F_0(q^2)$ at $Q^2=0$ 
can be obtained by adding in the non-relativistic approach 
exchange currents (pair term). At high $Q^2$, many statements 
made for the elastic case could be repeated here. They concern 
the ratio $F_0/F_1$, the comparison with non-relativistic calculations 
and the fall-off of the form factors in the ``point form'' approach.

\subsection{Other mass operators}
\label{subsec:34}

\noindent
Results presented in tables~(\ref{t10}-\ref{t30}) have been obtained with a 
wave function issued from a Coulomb-type potential (model v0). An important 
question is whether qualitative results obtained so far 
%as to the reliability of different approaches 
extend to other interaction models. For comparison, we used a 
wave function issued from a linear mass operator, eq.~(\ref{2m}), 
with an interaction given by eq.~(\ref{2n}). In this calculation, (model v1),
the coupling constant has been adjusted to reproduce the same binding energy 
as the one obtained with the Bethe-Salpeter equation and $\alpha=3$, giving 
$\alpha({\rm v}1)=1.775$. The difference in the couplings is simply due to the 
fact that 
the Bethe-Salpeter approach accounts for retardation effects which effectively 
decrease the strength of the interaction~\cite{Amghar:1999ii,Amghar:2000pc}. 
The model v1 does not reproduce the spectrum of the Wick-Cutkosky model as well
as the model v0.
The binding energy of the first radial excitation is $0.1320\,m$ instead 
of $0.1036\,m$. Therefore, a comparison of form factors  from this model with 
the exact ones is less instructive. However, 
a comparison with a non-relativistic type calculation, using eqs.~(\ref{2u}), 
may still be useful. 

No major qualitative difference with 
previous results is seen at small as well as high $Q^2$ (see table~\ref{t40}). 
Quantitative differences can be traced back to the interaction model itself. It 
provides a slightly more rapid decrease of the wave function in momentum space 
(roughly given by an extra factor $(e_k+m)/(2e_k)$ for small binding 
energy). This is a consequence 
of the semi-relativistic kinematics in eq.~(\ref{2k}) together with 
normalization factors in eq.~(\ref{2n}). As a  result, the wave function 
at the origin, $\psi_r(0)$, is smaller (see section~\ref{sect:4} 
for the role of this quantity). 

As a side remark, we notice 
that the asymptotic values for the form factors in the Coulombian model, v0, 
and the model v1 have not yet been reached at momentum transfers 
as large as $Q^2/m^2=100$. For v1, the value is too large by 
$\sim15\%$ ($\left.(Q^4/m^4)\,F(Q^2)\right|_{Q^2/m^2=100}\simeq 7.7$ instead 
of 6$.7$ asymptotically), while for v0 it is too low by $\sim10\%$ 
($\left.(Q^4/m^4)\,F(Q^2)\right|_{Q^2/m^2=100} \simeq 34$ instead of $38$). 
The ratio of the asymptotic values, $0.18$, is mainly due to the difference 
in the values of the wave functions at the origin ($0.24$), of the factor 
$(e_k+m)/(2e_k)$ at large $k$ ($0.5$) and of the coupling constants ($1.4$). 

\begin{table}[tb]
\caption{Elastic form factors  $F_0(q^2)$ and $F_1(q^2)$, 
calculated with wave functions
issued from the interaction models v1, given by eq.~(\protect\ref{2n}), and v2,
which includes the correction eq.~(\ref{2o}).
The couplings, $\alpha({\rm v}1)=1.775$ and $\alpha({\rm v}2)=1.327$, have been 
determined to reproduce the binding energy $E=0.4322\,m$  of the Bethe-Salpeter 
equation ($\alpha=3$). Results for the model v0 ($\alpha({\rm v}0)=1.241$) and
$B.S.$ are recalled for comparison.}
\label{t40}
\begin{ruledtabular}
\begin{tabular}{lccccc}
  $Q^2/m^2$     &   0.01   &   0.1  &  1.0   &  10.0 & 100.0 \\ [1.ex] \hline
 $N.R.$(v1)  &   &   &        &      &   \\ [0.ex] 
 $F_0=F_1$     & 0.995    & 0.953  & 0.640  &  0.665-01  & 0.774-03  \\ [1.ex]

 $P.F.$(v1)   &   &   &        &       &  \\ [0.ex] 
 $F_0$     & 0.710    & 0.651  & 0.303  &  0.596-02  & 0.281-05  \\  [0.ex]    
 $F_1$     & 0.992    & 0.924  & 0.493  &  0.205-01  & 0.464-04  \\  [1.ex]    
\hline  
    
 $N.R.$(v2)  &   &   &        &      &   \\ [0.ex] 
 $F_0=F_1$     & 0.996    & 0.966  & 0.727  &  0.132  & 0.279-02  \\ [1.ex]

 $P.F.$(v2)   &   &   &        &       &  \\ [0.ex] 
 $F_0$     & 0.689    & 0.645  & 0.358  &  0.140-01  & 0.117-04  \\  [0.ex]    
 $F_1$     & 0.994    & 0.946  & 0.603  &  0.494-01  & 0.177-03  \\  [1.ex]    
\hline  

 $N.R.$ (v0) &   &   &        &      &   \\ [0.ex] 
 $F_0=F_1$     & 0.997    & 0.968  & 0.740  &  0.146  & 0.338-02  \\ [1.ex]

 $P.F.$(v0)   &   &   &        &       &  \\ [0.ex] 
 $F_0$     & 0.682    & 0.641  & 0.366  &  0.16-01  & 0.15-04  \\  [0.ex]    
 $F_1$     & 0.995    & 0.949  & 0.621  &  0.56-01  & 0.23-03  \\  [1.ex] 
\hline  
 
 $B.S.$   &   &   &        &       &  \\ [0.ex] 
 $F_0$      & 1.123    & 1.080  & 0.767  &  0.132  & 0.394-02  \\ [0.ex] 
 $F_1$      & 0.996    & 0.962  & 0.705  &  0.139  & 0.503-02  \\ [1.ex] 

\end{tabular}
\end{ruledtabular}
\end{table}

The origin of the above quantitative differences can be checked 
by using a model more in the spirit of the mass operator of eq.~(\ref{2k}), 
like the one incorporating corrections to the interaction as given 
by eq.~(\ref{2o}). In this model, denoted v2, the coupling is again 
fitted to the binding energy obtained with the Bethe-Salpeter equation, 
giving $\alpha({\rm v}2)=1.327$, which is quite close to the one 
for the Coulombian model $\alpha({\rm v}0)=1.241$. The corresponding 
results for the form factors are also shown in table~\ref{t40}. As 
expected, they get closer to those quoted as non-relativistic ones 
or to the exact ones. However, the form of the interaction prevents one 
from determining in an easy way the expression of the associated two-body 
currents. Keeping in mind that this extra set of results ensured a continuous 
transition between the results obtained with different interaction models, 
we will only consider in sect.~\ref{sect:5} the simplest case, v1, for which 
two-body currents can also be derived without too much difficulty, 
while remaining close to realistic ones.

%%%%%%%%%%%%%%%%%%%%%%%%%%%%%%444444444444444444444%%%%%%%%%%%%%%%%%%%%%%%%%%%%

\section{Remarks concerning form factors at low and high  $Q^2$} 
\label{sect:4}

\noindent
Previous ``point form'' results obtained in the single-particle current
approximation were found to significantly depart from the ``exact'' 
ones. We analyze on general grounds the role of further contributions 
to the current in correcting form factors,  successively at low 
and high $Q^2$. In one case, they concern low energy theorems 
and consistency properties, while in the other, they involve 
the Born amplitude.

%%%%%%%%%%%%%%%%%%%%%%%%%%%%%%%%%4.1
\subsection{Analysis of results at low $Q^2$}
\label{subsec:41}

\noindent
A detailed examination of the ``point form'' calculation of form factors shows 
that a large part of the difference with the non-relativistic calculation 
can be traced back to the relativistic kinematical boost effect~\cite{Wagenbrunn:2000es}.  
The quantity  $p_z \pm \frac{Q}{4}$, which appears in the non-relativistic 
Breit-frame expression of the form factor for the ground state 
for instance (Coulombian case):
\begin{eqnarray}
\label{4a}
\nonumber
\lefteqn{
I^{N.R.} \propto \int d \vec{p} 
\left(\kappa^2+p_x^2+p_y^2+(p_z-\frac{Q}{4})^2  
\right)^{-2}} \\ && \qquad \qquad 
\left(\kappa^{2}+p_x^2+p_y^2+(p_z+\frac{Q}{4})^2  
\right)^{-2},
\end{eqnarray}
is replaced by  $\frac{p_z \pm v\,e_p}{\sqrt{1-v^2}}$
 to give:
\begin{eqnarray}
\label{4b}
\nonumber
\lefteqn{
I^{P.F.} \propto \int d \vec{p} 
\left(\kappa^2+p_x^2+p_y^2+(\frac{p_z-v\,e_p}{\sqrt{1-v^2}})^2  
\right)^{-2}} \\ && \qquad \qquad 
\left(\kappa^{2}+p_x^2+p_y^2+(\frac{p_z+v\,e_p}{\sqrt{1-v^2}})^2  
\right)^{-2}.
\end{eqnarray}
To emphasize differences with the non-relativistic expression, we rewrite 
the term $(p_z \pm v\,e_p)/\sqrt{1-v^2}$ as:
\begin{equation}
\frac{p_z \pm v\,e_p}{\sqrt{1-v^2}}=\frac{p_z}{\sqrt{1-v^2}} \pm
\frac{Q}{4}\, \frac{2\sqrt{m^2+p^2}}{M}.\label{4c} 
\end{equation}
This differs from the non-relativistic expression in two ways:
the factor multiplying $Q/4$, $(2\sqrt{m^2+p^2})/M$, and the 
factor multiplying $p_z$, $1/\sqrt{1-v^2}$. They are successively 
analyzed in the following.

\subsubsection{The factor $(2\sqrt{m^2+p^2})/M$}

\noindent
The extra factor
$(2\sqrt{m^2+p^2})/M$ which multiplies the quantity $Q/4$ in eq.~(\ref{4c}) 
is always larger than one, both because the numerator, 
$2\,\sqrt{m^2+p^2}$, is larger than $2\,m$ and that the total mass 
of the system, $M$, at the denominator is smaller than the same quantity. 
In some sense, the momentum transfer $Q$ entering the non-relativistic 
calculation should be replaced by an effective one, which is larger, 
leading to an effective scaling of the electromagnetic properties 
similar to that one found in ref.~\cite{Wagenbrunn:2000es}. 
The effect is especially 
large when the average momentum of particles composing the system under 
consideration is large, as it is in the nucleon wave functions employed 
in this last reference, or when the total mass of the system goes to zero.

Many examples indicate that one should be cautious about effects 
involving a factor like $(2\,\sqrt{m^2+p^2})/M$. Often, the kinetic 
energy $e_p$ combines with the potential energy $V$ to give 
the total energy, $M$. In such a case, results based on the above 
expression, or a similar one, could be affected.

$\bullet$ In a nuclear mean-field approach intending to incorporate 
relativistic effects, a perturbative calculation of a matrix element 
of the current can be performed by retaining positive energy spinors,
$ \propto \left(1, \frac{\sigma \cdot p}{e_p+m}\right) | \chi \rangle$, 
to describe spin-1/2 particles. Incorporating the  contribution of mesonic 
exchange currents (pair term essentially), it is found that 
it combines with the single-particle one in such a way that 
the final result involves the total energy $\epsilon$ in place 
of $e_p$ at the denominator of the small component. This is what 
one would obtain using directly solutions of the Dirac equation 
with the full interaction instead of free particle ones. The kinetic 
energy term has thus been completed with an interaction energy 
to provide the total binding energy.

$\bullet$ In the nuclear Dirac phenomenology, large effective mass effects 
have been found. The orbital part of the magnetic moment in nuclei appeared 
enhanced by a factor $m/m^*$ in first relativistic mean field approaches. 
It took some time to realize where the problem came from and how 
to solve it: by considering general conditions to be fulfilled from symmetry 
arguments, such as the relation of the vector current to the time component 
(four-vector, iso-scalar part) given by Galilean- or Lorentz covariance. 
The total current, for instance, is given by the quantity 
$\sum_i \vec{p}_i/m$ rather than $\sum_i \vec{p}_i/m^{*}$. 
The enhancement of the iso-scalar magnetic 
moments then vanishes, which is achieved only from current conservation. 

$\bullet$ When verifying the current conservation in the non-relativistic 
case, or what accounts for it in a more general case, the divergence 
of the vector current, $ \vec{q} \ccdot  \langle\vec{J}_{I.A.}\rangle$, 
involves the difference of the kinetic energies corresponding to the 
initial and final states, multiplied by the time component of the 
current $\langle J^0\rangle$. When the many-body part of the current 
is added, an extra term equal to the difference of the potentials 
relative to the initial and final states also appears. These ones 
combine with the kinetic energies to provide the difference 
in the total energies, $\epsilon_i-\epsilon_f$, times $\langle J^0\rangle$, 
allowing current conservation to be fulfilled.  An equation of the form 
``$\sum e_p+\sum V=\epsilon$", has to be used. An example of how 
current conservation is fulfilled is worked out in some detail in 
sect.~\ref{sect:5}. A quite similar argument underlies the Siegert theorem. 
In the simplest case of an $E1$ electric transition in the long range 
limit, it states that the sum of the one-body and many-body contributions 
to $\vec{\langle J \rangle}$, which respectively have a kinetic and 
an interaction character, is equal to the matrix element of the dipole 
operator times the difference of the energies of the initial and 
final states, $\langle \sum_j\vec{r}_j \rangle\,(\epsilon_i-\epsilon_f)$. 

$\bullet$ In calculating the current of a two-body system made 
of two scalar constituents with $l=0$ and total momentum $\vec{P}$, 
one gets in impulse approximation the contribution
\begin{equation}
\vec{J} \propto \frac{\vec{p}_1}{e_{p_1}} + \frac{\vec{p}_2}{e_{p_2}}. 
\label{4d}
\end{equation}
This one generally differs from what is expected from symmetry arguments, 
the Lorentz covariance in the present case:
\begin{equation}
\vec{J} \propto 2\,\frac{\vec{P}}{\sqrt{M^2+P^2}}.
\label{4e}
\end{equation}
To recover  the correct result, extra two-body contributions have 
to be considered. Simplifying somewhat the argument, they have to provide 
an interaction term which will add to the kinetic energy term $e_p$ 
appearing in the denominator of the single-particle current of 
eq.~(\ref{4d}), to give the total mass of the system, $M$, 
in the limit of small values of $\vec{P}$. Part of the necessary 
two-body contribution is due to the standard pair current. 

$\bullet$ The asymptotic normalization of the wave function 
at large distances, $\psi(r)_{r\rightarrow \infty}=A_S \exp(-\kappa r)/r$, 
is an observable quantity (for the deuteron for instance), contrary 
to the wave function itself. In momentum space, this normalization 
is related to the pole of the wave function at $k=i\kappa$. For the 
scattering problem, this pole involves the scattering phase shift. 
In calculations of the wave functions in the light-front approach, 
new components that depend on the light-front orientation $\vec{n}$
appear. At the above pole, this dependence should vanish, since the 
light-front orientation is arbitrary. This constraint is not fulfilled 
in a perturbative calculation~\cite{Carbonell:1995yi}. To remedy this situation, 
higher order terms in the interaction have to be considered. This 
was shown in ref.~\cite{Desplanques:1995} for the deuteron case and at the lowest 
order in the interaction. The dominant $\vec{n}$ - dependent component 
in the perturbative calculation, produced by the $\pi$-exchange 
in the pseudo-scalar coupling, reads:
\begin{eqnarray}
\label{4f}
\nonumber
\lefteqn{
f_5^0[\vec{k}\times\vec{n}]/k = -\frac{3\,g^2_{\pi NN}}{4\sqrt{3}\, m^2} \,
\frac{m}{k^2+\kappa^2}} \\ && 
\int\frac{d\vec{k}'}{(2\pi)^3} \,
\frac{(\vec{k}-\vec{k}')\times \vec{n}}{\mu^2+(\vec{k}-\vec{k}')^2} \, 
\frac{(k^2-k'^2)}{m}\, u_S(\vec{k}'),
\end{eqnarray}
where $u_S(\vec{k}')$ represents the deuteron S-wave. It can be easily 
checked that the pole at $k=i\kappa$ does not vanish. However, 
when higher order effects in the interaction are accounted for, a new 
contribution appears and adds to the kinetic energy term $k'^2/m$ 
under the integral, to provide, up to the mass 
factor $m$, the total interaction, $k'^2/m +V$. Using the equation 
fulfilled by the component $u_S(\vec{k}')$ at the lowest order in the 
interaction, one can replace the total interaction by $-\kappa^2/m$. The 
factor $(k^2-k'^2)/m$ under the integral thus
becomes  $(k^2+\kappa^2)/m$ and
cancels the front factor $m/(k^2+\kappa^2)$. The pole of eq.~(\ref{4f}) 
at $k=i\kappa$ now vanishes. This example shows how an interaction term 
adds to a kinetic energy term to provide the total energy, allowing 
one to get consistent results.

Assuming that the observation made above on several examples also works in the
present case, 
one should add in eq.~(\ref{4c}) an interaction term $V$ so that 
the factor multiplying $Q/4$ now reads $(2\sqrt{m^2+p^2}+V)/M$. 
Taking into account that the numerator acting on a wave function is 
nothing but $M$, the factor may be equal to one, as the analysis 
of the triangle Feynman diagram  tends to show~\cite{Desplanques:2001ze}. 
This immediately removes the scaling of electromagnetic properties 
mentioned at the beginning of this subsection, making the results 
closer to the exact and the non-relativistic ones.

\subsubsection{The factor $1/\sqrt{1-v^2}$}

\noindent
Another consequence of the relativistic boost is the appearance of 
the factor $1/\sqrt{1-v^2}$ multiplying $p_z$. As outlined in
appendix~\ref{app:c}, a simple change of variable 
allows one to remove it from the integrand in eq.~(\ref{4b}) and to 
factor out the quantity $\sqrt{1-v^2}$. Up to this factor, the integral, 
eq.~(\ref{4b}), then becomes identical to its non-relativistic limit, 
eq.~(\ref{4a}). Notice that the result involves the very dependence 
of the wave function on the momentum. To evidence the ambiguous 
character of the above change of variables, it suffices to replace the 
factor at the denominator of the wave function employed in eq.~(\ref{4a}), 
$\kappa^2+\vec{p}^{\,2}$, by  the equivalent $\kappa^2+e_p^2-m^2$. 
If $e_p$ happens to combine with an interaction term, as discussed in the
previous subsection, to give an overall mass term, there would be no more
momentum dependence and the result would be quite different. 
In this case however, the replacement of the genuine momentum dependence 
into an energy dependence has no theoretical foundation but this 
may be different for other factors entering the calculation. 

The above developments are schematic ones. They are mainly intended 
to show in a few cases how relativistic effects (kinematical boost 
and pair current for instance) combine to give a total result 
in agreement with expectations from some symmetry (current conservation 
in most examples considered here, rotational invariance, ...). 
The idea behind getting together those contributions is that, when 
a boost is made, not only the kinetic energy which enters the total mass 
of a system is boosted, but also the potential energy part. In view 
of the various examples presented in this subsection, one should 
therefore look with much caution at the present results in the ``point form'' 
approach, especially those at small $Q^2$ like the scaling of some 
properties with the inverse of the total mass of the system, $M$. 
  
%%%%%%%%%%%%%%%%%%%%%%%%%%%%4.2
\subsection{Analysis of results at high $Q^2$}
\label{subsec:42}

\begin{figure}[tb]
\begin{center}
\mbox{
\includegraphics[width=0.4\columnwidth]{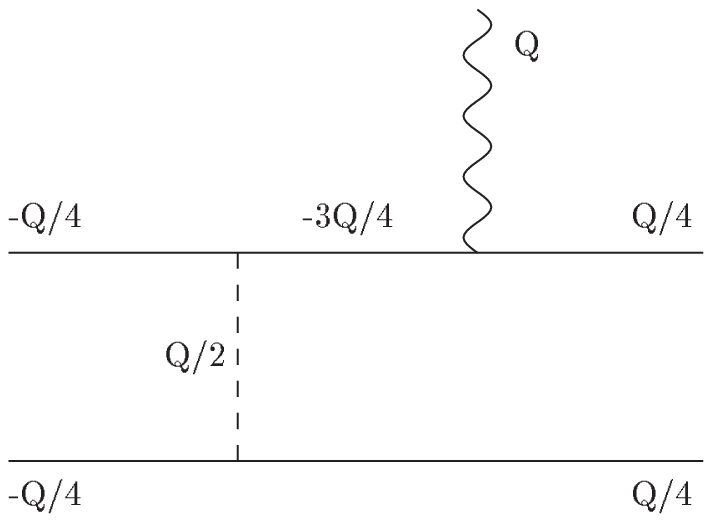} \hspace{2em}
\includegraphics[width=0.4\columnwidth]{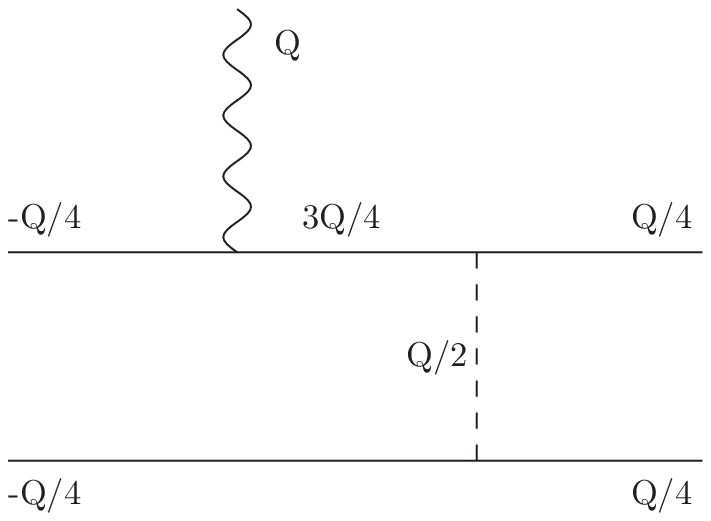}}
\end{center}
\caption{Virtual scalar particle or photon absorption on a two-body 
system in Born approximation. The kinematical definitions refer to the 
Breit-frame. They can be used for both the Feynman diagram and the 
non-relativistic (or instant-form) approach where the 3-momenta are 
conserved at all vertices, but not in the ``point form'' approach where 
a different conservation law holds.\label{fig2}}
\end{figure}

\noindent
At high $Q^2$ it is expected that form factors are dominated by the 
contribution of the full Born amplitude represented by the Feynman 
diagram shown in Fig.~\ref{fig2}. It is on this basis that Alabiso 
and Schierholz made predictions for form factors in the asymptotic 
domain~\cite{Alabiso:1974sg}. All calculations employing wave functions 
obtained from some equation together with some interaction provide 
a contribution to the full diagram shown in this figure. Using 
a perturbative-type approach, this contribution can be calculated. 
By comparing it to the full diagram, one can 
determine how it does in predicting the high $Q^2$ behavior of 
form factors with respect to the underlying theory. What is missing 
may be incorporated in two-body currents. In this subsection, we 
analyze the contribution of each approach in the Born approximation. 
This follows lines developed in various papers, 
especially in ref.~\cite{Desplanques:2000ev}. 

Beginning with the non-relativistic calculation for the ground state, 
it is found that the form factors at high $Q^2$ can be expressed 
as the product of the Born amplitude times the squared wave function 
at the origin (in configuration space):
\begin{eqnarray}
\label{4g}
\nonumber
\lefteqn{
\left.F_0(q^2)\right|_{Q^2\rightarrow \infty}=  
\left.F_1(q^2)\right|_{Q^2\rightarrow \infty}} \\
\nonumber && \qquad= 
2 \, \psi^r(0) \, \left.\psi^p(p=Q/2)\right|_{Q^2\rightarrow \infty} \\
& &\qquad\qquad=  2 \, Born \,\left.(1\,diagr.)\right|_{Q^2\rightarrow \infty} 
\,\psi_r^2(0). 
\end{eqnarray}
The wave function at the origin, $\psi_r(0)$, should be determined 
numerically. In the Coulombian case referred to in tables~\ref{t10} 
and~\ref{t20}, 
it is given by $\psi_r^2(0)=\kappa^3/\pi$. As for the 
Born amplitude in the non-relativistic case, it reads:
\begin{equation}
Born\,\left.(1\,diagr., \,N.R.)\right|_{Q^2\rightarrow 
\infty}=\frac{g^2}{\mu^2+Q^2/4} \, \frac{4\,m}{Q^2}.
\label{4h}
\end{equation}
It corresponds to the product of a term involving the interaction 
and the propagator for the two constituent particles, both being 
calculated in any frame but consistently with Galilean invariance. 
In order to easily identify the interaction term, the mass of the 
exchanged boson, $\mu$, is written explicitly even though it is taken 
as zero in actual calculations performed later on.
From the above result, it is immediately seen that the form factors 
scale like $1/Q^4$, factors $1/Q^2$ being contributed separately 
by the boson and the constituent propagators in Fig.~\ref{fig2}, 
in agreement with the standard counting rules for determining the 
high $Q^2$ behavior of form factors. This roughly explains 
the behavior of form factors evidenced by the ``exact'' results 
shown in tables~\ref{t10} and~\ref{t20} and, of course, 
in the non-relativistic case.  

The above result can be refined by considering the full Feynman diagrams
shown in Fig.~\ref{fig2}. These ones can be split into two terms where 
the intermediate constituent propagates with positive and negative 
energies. The details may depend on the formalism or on the frame. 
The expressions have a form similar to eq.~(\ref{4g}), except that 
scalar and charge form factors now differ and involve corrections 
of relativistic order:
\begin{eqnarray}
\label{42c} 
\left.F_0(q^2)\right|_{Q^2\rightarrow \infty} & = & 2 \, 
\left.(B.A._0)\right|_{Q^2\rightarrow \infty} 
\frac{N}{4m} \tilde{\psi}_r^2(0), \nonumber\\
\left.F_1(q^2)\right|_{Q^2\rightarrow \infty} & = & 2 \, 
\left.(B.A._1)\right|_{Q^2\rightarrow \infty} 
\frac{N}{E_i+E_f} \tilde{\psi}_r^2(0), \nonumber\\
& {\rm with}  & \tilde{\psi}_r(0) = 
\int \frac{d\vec{p}}{(2\pi)^3} \frac{m}{e_p} \phi(\vec{p}\,).
\end{eqnarray}
The normalization constant $N$ is defined in eq.~(\ref{appd2}) and the factors
$4m$ and $E_i+E_f$ cancel a corresponding factor in the Born amplitude such that
we recover the definition of the form factors, see eq.~(\ref{2a}). For the 
instant-form formalism and in the Breit-frame, the Born amplitudes 
in the asymptotic limit read:
\begin{eqnarray}
\label{4j}
\nonumber 
B.A._0 &=& \frac{g^2}{\mu^2+Q^2/4} \,\, \frac{1}{2e_{3Q/4}} \\
&& \nonumber \times
\left(\frac{2m}{e_{3Q/4}-e_{Q/4}} + \frac{2m}{e_{3Q/4}+e_{Q/4}}\right)  \\
 &=& \frac{g^2}{\mu^2+Q^2/4} \,\, \frac{2m}{e^2_{3Q/4}-e^2_{Q/4}}, 
\end{eqnarray}
\begin{eqnarray}
\label{4k} 
\nonumber
B.A._1 &=&\frac{g^2}{\mu^2+Q^2/4} \,\, \frac{1}{2e_{3Q/4}} \\ 
&& \nonumber \times
\left(\frac{e_{3Q/4}+e_{Q/4}}{e_{3Q/4}-e_{Q/4}} + 
\frac{-e_{3Q/4}+e_{Q/4}}{e_{3Q/4}+e_{Q/4}}\right)  \\
 &=& \frac{g^2}{\mu^2+Q^2/4}\,\, \frac{2e_{Q/4}}{e^2_{3Q/4}-e^2_{Q/4}}  . 
\end{eqnarray}
We omit normalization factors $m/e$ in the above equations. As mentioned 
previously, these ones have to be accounted for when the corresponding 
contributions to form factors are calculated, see eq.~(\ref{42c}). In the 
non-\-relativistic limit, the positive energy part of the constituent 
propagator allows one to recover the non-relativistic result given 
by eq.~(\ref{4h}). At very high $Q^2$ the total form factors are 
identical to the non-relativistic ones (the normalization factor 
$\frac{N}{4m} \tilde{\psi}_r^2(0)$, of the order of 1, put apart), 
but evidence a difference in the relative contributions of the 
different parts of the constituent propagator:
\begin{eqnarray}
\label{4l}
\nonumber
\lefteqn{
 \left.F_0(q^2)\right|_{Q^2\rightarrow \infty} = } \\ \nonumber &&
2\, \frac{g^2}{\mu^2+Q^2/4} \,\, \frac{4\,m}{Q^2} \, ( \frac{2}{3} + 
\frac{1}{3}) \,  \frac{N}{4m} \tilde{\psi}_r^2(0), \\ \nonumber
\lefteqn{
\left.F_1(q^2)\right|_{Q^2\rightarrow \infty} = } \\ &&
2\, \frac{g^2}{\mu^2+Q^2/4} \,\, \frac{4\,m}{Q^2} \, ( \frac{4}{3} - 
\frac{1}{3}) 
\,  \frac{N}{4m} \tilde{\psi}_r^2(0) .
\end{eqnarray}
The impulse-approximation calculation of asymptotic form factors 
in the ``point form'' approach could be performed along the above lines, 
with some modifications concerning the kinematics. The structure 
of the result is similar to eq.~(\ref{4h}),
\begin{equation}
Born\,\left.(1\,diagr., \,P.F.)\right|_{Q^2\rightarrow 
\infty}=\frac{g^2}{\mu^2+\tilde{Q}^2/4} \,\frac{4\,m}{\tilde{Q}^2},
\label{4m}
\end{equation}
but the momentum transfer $Q^2$ is replaced by 
$\tilde{Q}^2=Q^2\,\left[1+ Q^2/(4M^2)\right]$, recovering what 
was obtained in ref.~\cite{Allen:2000ge}. For the electromagnetic probe, 
an extra factor $(1+v^2)/(1-v^2)$  has 
to be added, in relation with the different coupling to the 
external probe. This is sufficient to explain the overall behavior 
of form factors shown in tables~\ref{t10} and~\ref{t20}. Careful examination 
however indicates that there are other corrections than the one given in 
eq.~(\ref{4m}) that give contributions with a log character.

In practice, depending on precise details in the formalism, 
the various contributions in eqs.~(\ref{4j},\ref{4k}) may
have a different weight but the overall result, eq.~(\ref{4l}), 
should be recovered. This evidently applies to the ``point form'' results. 
In this case however, the situation is somewhat different because 
one has to completely rely on two-body currents to get the right 
asymptotic power-law behavior of form factors.

%%%%%%%%%%%%%%%%%%%%%%%%%%5555555555555555555555%%%%%%%%%%%%%%
\section{Two-body currents: expressions and results}
\label{sect:5}

\noindent
We consider here specific models for two-body currents. They 
are constructed via the requirement of current conservation and reproducing 
the Born amplitude. Methods allowing one to get these currents as well as 
their limitations are known~\cite{Gross:1987bu,Coester:1994wp}. 
They are adapted to our purpose in the case of the relativized model, v1. 
After discussing a norm correction in relation with the ratio $F_0/F_1$, 
which in some sense also involves two-body currents, a presentation 
of numerical results is made.

 %%%%%%%%%%%%%%%%%%%%%%%%%%%%%5.1
\subsection{Expressions of two-body currents}
\label{subsec:51}

\noindent
We already mentioned that two-body currents are required in most 
approaches to satisfy current conservation and could also be required 
to get the right Born amplitude. This second property may not be 
related to the first one, current conservation holding up to terms  
that are gauge invariant by themselves. 

There are methods that allow one to derive contributions that restore 
current conservation but the result may not be quite satisfying, 
either because it misses the Born amplitude or because it is very 
cumbersome. Here, we favor the high momentum transfer region, and 
therefore the  Born amplitude, and simplicity.

%%%%%%%%%%%%%%
\subsubsection{Two-body currents motivated by current conservation in the 
non-relativistic case}

\noindent
We begin with an interaction model that represents an extension 
of the model v1 of eq.~(\ref{2n}) to any frame (in the instant form):
\begin{eqnarray}
\label{5a}
\nonumber
\lefteqn{
V_{int}(\vec{p}_1,\vec{p}_2,\vec{p}_1\!',\vec{p}_2\!')= -
\delta(\vec{p}_1+\vec{p}_2 -\vec{p}_1\!' - \vec{p}_2\!')} \\ && \qquad
\sqrt{\frac{m}{e_{p_1}} \, \frac{m}{e_{p_2}} } \,\,
\frac{g^2}{ \mu^2+(\vec{p}_2-\vec{p}_2\!')^2}\,\,
\sqrt{ \frac{m}{e_{p_1'}} \, \frac{m}{e_{p_2'}} }.
\end{eqnarray}
The corresponding equation to be solved in principle generalizes 
eq.~(\ref{2k}):
\begin{eqnarray}
\label{5b}
\nonumber
\lefteqn{
(E-e_{p_1}-e_{p_2})\,\Phi(\vec{p}_1,\vec{p}_2)=} \\ && 
\int\!\!\!\!\int 
\frac{d\vec{p}_1\!'}{(2\pi)^3}\frac{d\vec{p}_2\!'}{(2\pi)^3}
V_{int}(\vec{p}_1,\vec{p}_2,\vec{p}_1\!',\vec{p}_2\!')\,
\Phi(\vec{p}_1\!',\vec{p}_2\!').
\end{eqnarray}
Though the set of eqs.~(\ref{5a},\ref{5b}) is not the one 
we will use for actual calculations, it offers the great advantage, 
due to its close relation to a field-theory approach, that currents 
take a relatively simple form, allowing one to illustrate some 
of the peculiarities relative to their derivation. It is noticed 
that the solutions for the mass $M$ only make sense 
in the non-relativistic limit as they in principle depend on the 
total momentum. A complete interaction kernel would be required 
to make the solutions meaningful so that to fulfill relativistic 
covariance. This is not however necessary for the following developments.

The single-particle current stems from the same field-theory that 
motivates the above interactions. It  is given for particle 1 by: 
\begin{eqnarray}
J^0_{I.A.}=  \frac{e_{p_1}+e_{p_1'}}{2 \sqrt{e_{p_1'}\,e_{p_1}}}\, 
\delta(\vec{q}+\vec{p}_1-\vec{p}_1\!')\, ,\nonumber \\
\vec{J}_{I.A.}=\frac{\vec{p}_1+\vec{p}_1\!'}{2 \sqrt{e_{p_1'}\,e_{p_1}}}\, 
\delta(\vec{q}+\vec{p}_1-\vec{p}_1\!').
\label{5c}
\end{eqnarray}
The above current provides a non-zero four-divergence which, using 
eq.~(\ref{5b}), can be written as:
\begin{eqnarray}
\label{5d}
q_{\mu} \ccdot J^{\mu}_{I.A.}&=&
\sqrt{\frac{m}{e_{p_1'}} \, \frac{m}{e_{p_2'}} } \,\,\frac{g^2}{ 
\mu^2+(\vec{p}_2-\vec{p}_2\!')^2}\,\,
\sqrt{ \frac{m}{e_{p_1}} \, \frac{m}{e_{p_2}} }  \nonumber \\ && \nonumber 
\times\left( \frac{e_{p_1'}-e_{p_1'-q} }{2\,e_{p_1'-q}} + 
      \frac{e_{p_1+q}-e_{p_1} }{2\,e_{p_1+q}}\right)  \\ &&
      \qquad \times \,\delta(\vec{q}+\vec{p}_1+\vec{p}_2 -\vec{p}_1\!' - \vec{p}_2\!') .
\end{eqnarray}
This has to be canceled by the four-divergence of a two-body 
current~\cite{Coester:1994wp}, which can 
be easily obtained in the present case:
\begin{eqnarray}
\label{5e}
\nonumber
\lefteqn{
\vec{J}_{int}(\vec{q},\vec{p}_1,\vec{p}_2, 
\vec{p}_1\!',\vec{p}_2\!')= } \\ && \nonumber 
\sqrt{\frac{m}{e_{p_1'}} \, \frac{m}{e_{p_2'}} } \,\,\frac{g^2}{ 
\mu^2+(\vec{p}_2-\vec{p}_2\!')^2}\,\,
\sqrt{ \frac{m}{e_{p_1}} \, \frac{m}{e_{p_2}} } 
 \\ && \nonumber \times
\left(\frac{2\,\vec{p}_1\!'-\vec{q}}{2\,e_{p_1'-q}\,(e_{p_1'}+e_{p_1'-q})} + 
 \frac{2\,\vec{p}_1+\vec{q}}{2\,e_{p_1+q}\,(e_{p_1}+e_{p_1+q})}
\right) \\ && \qquad 
\times\,\delta(\vec{q}+\vec{p}_1+\vec{p}_2 -\vec{p}_1\!' - \vec{p}_2\!').
\end{eqnarray}
Notice that 
one can easily recognize in this equation the structure of a pair term. 
By construction, it allows one to satisfy current conservation. 
When checking this property, it is found that the one- and two 
body-currents provide the following contributions (in the operatorial sense):
\begin{eqnarray}
\vec{q}  \ccdot   \vec{J}_{I.A.}= 
(e_{p_1}\,O(1)-O(1)\,e_{p_1}), \nonumber\\
\vec{q}  \ccdot   \vec{J}_{int}= 
(V\,O(1)-O(1)\,V), \nonumber
\end{eqnarray}
where $O(1)$ just represents the charge operator. After adding a contribution 
$(e_{p_2}\,O(1)-O(1)\,e_{p_2})$, which is zero, they combine to give:
\begin{equation}
\vec{q}  \ccdot   (\vec{J}_{I.A.}+\vec{J}_{int})= 
H\,O(1)-O(1)\,H,
\label{5f}
\end{equation}
The last equality, taken between eigen-states of the Hamiltonian, 
is the product of the energy transfer, $q_0$, times the charge operator. 
This provides another illustration of how contributions involving 
the kinetic energy and the potential energy separately add together 
to give the total energy of the system. We notice that the above 
two-body current does not contain any $1/q^2$ factor as some
recipe enforcing current conservation,
\begin{equation}
J^{\mu} \rightarrow J^{\mu} - q^{\mu} \, J \ccdot  q \, /  q^2,
\label{5g}
\end{equation}
would suppose~\cite{Allen:2000ge}. Notice that 
in the small $q$ limit, the two-body current obtained in eq.~(\ref{5e}) 
has the schematic form $-\partial_{q_{\mu}} (``J \ccdot  q")$, 
where $``J \ccdot  q"$ is given by the right-hand side 
of eq.~(\ref{5d}). It therefore significantly differs 
from the term introduced in eq.~(\ref{5g}) and, evidently, it has 
no singular character in the limit $q \rightarrow 0$. 

%%%%%%%%%%%%
\subsubsection{Two-body currents motivated by the  Born amplitude in the 
non-relativistic case}

\noindent
The contribution derived above is not sufficient to recover 
the Born amplitude. Starting from this requirement, another two-body current 
is obtained, which, underlying the theoretical model under 
consideration, also contributes to the time component of the current, 
contrary to the interaction term. The extra term to be added 
to eq.~(\ref{5e}) is self-gauge invariant. To emphasize this 
feature, it is written in a way where this property is readily 
satisfied, i.e. by introducing the photon polarization $\epsilon^{\mu}$:
\begin{widetext}
\begin{eqnarray}
\label{5h}
\nonumber
\lefteqn{
\left(\epsilon^0\,  J^0_{\Delta B}- 
\vec{\epsilon}\cdot \vec{J}_{\Delta B}\right)(\vec{q},\vec{p}_1,\vec{p}_2, 
\vec{p}_1\!',\vec{p}_2\!')= } \\ && \nonumber
\sqrt{\frac{m}{e_{p_1'}} \, \frac{m}{e_{p_2'}} } \,\,\frac{g^2}{ 
\mu^2+(\vec{p}_2-\vec{p}_2\!')^2}\,\,
\sqrt{ \frac{m}{e_{p_1}} \, \frac{m}{e_{p_2}} } 
\delta(\vec{q}+\vec{p}_1+\vec{p}_2 -\vec{p}_1\!' - \vec{p}_2\!') 
\left(\epsilon^0 \,\vec{q}-\vec{\epsilon}\, q^0\right)\\ && 
\cdot 
\left(\frac{2\,\vec{p}_1\!'-\vec{q}}{2\,e_{p_1'-q}\,(e_{p_1'}+e_{p_1'-q})\,
(e_{p_1}+ e_{p_2}- e_{p_2'}+e_{p_1'-q})} -
\frac{2\,\vec{p}_1+\vec{q}}{2\,e_{p_1+q}\,(e_{p_1}+e_{p_1+q})\,
(e_{p_1'}+e_{p_2'}-e_{p_2}+e_{p_1+q})}\right).
\end{eqnarray}
While deriving these currents, off-shell effects have been neglected, 
which amount to corrections of the order $g^4$. This is done consistently 
with neglecting higher order contributions in the amplitude. Notice 
that the subscript ${\Delta B}$ in eq.~(\ref{5h}) and later on 
does not refer to the Born amplitude itself. It represents 
the contribution that has to be added to the impulse approximation plus 
interaction terms in order to recover the Born amplitude.

While current conservation tells us nothing about two-body contributions 
in the case of a scalar probe, requiring that the Born amplitude provided 
by the Feynman diagram of Fig.~\ref{fig2} be reproduced imposes to consider 
further terms. These ones, for an interaction of particle $1$ with 
the external probe, are given by:
\begin{eqnarray}
\label{5i}
\nonumber
\lefteqn{
 S_{\Delta B}(\vec{q},\vec{p}_1,\vec{p}_2, \vec{p}_1\!',\vec{p}_2\!')= 
\sqrt{\frac{m}{e_{p_1'}} \, \frac{m}{e_{p_2'}} } \,\,\frac{g^2}{ 
\mu^2+(\vec{p}_2-\vec{p}_2\!')^2}\,\,
\sqrt{ \frac{m}{e_{p_1}} \, \frac{m}{e_{p_2}} }}  \\ && \times
\left(\frac{2\,m}{2\,e_{p_1'-q}\,
(e_{p_1}+ e_{p_2}- e_{p_2'}+e_{p_1'-q})} 
 + \frac{2\,m}{2\,e_{p_1+q}\, 
(e_{p_1'}+e_{p_2'}-e_{p_2}+e_{p_1+q})}
\right)\, \delta(\vec{q}+\vec{p}_1+\vec{p}_2 -\vec{p}_1\!' - \vec{p}_2\!').
\end{eqnarray}
%\end{widetext}
%
It is noticed, not surprisingly, that the contributions from 
eqs.~(\ref{5h},\ref{5i}) to form factors identify to the second term 
at the r.h.s. of eqs.~(\ref{4j},\ref{4k}) in the same limit. 
This is due for a part to the instant-form formalism which underlies 
both expressions.

%%%%%%%%%%%%%%
\subsubsection{Two-body currents motivated by current conservation in 
 the ``point form'' approach}

\noindent
The first step in deriving two-body currents for calculating form factors 
in the ``point form'' is the definition of the interaction corresponding 
to the interaction mentioned previously. Its invariant form involves 
the four-velocity $\lambda^{\mu}$ of the system which we are interested in:
\begin{equation}
\label{5j}
V_{int}(p_1,p_2,p'_1,p'_2)=
-\sqrt{\frac{m}{\lambda \ccdot p_1}\,\frac{m}{\lambda \ccdot p_2}} \,\,
\int d\eta \frac{g^2\, \delta^4(p_1+p_2 -p_1' - p_2'-\lambda \eta) }{ 
\mu^2-(p_2-p'_2)^2+(\lambda  \ccdot   (p_2-p'_2) 
)^2 }\,\,
\sqrt{ \frac{m}{ \lambda  \ccdot   p'_1} \, 
\frac{m}{\lambda  \ccdot   p'_2 } },
\end{equation}
with $ \lambda  \ccdot   (p_1-p_2)=\lambda \ccdot   (p'_1-p'_2)=0$. 
Equation~(\ref{5b}) then reads:
\begin{eqnarray}
\label{5k}
\nonumber
\lefteqn{
(M - \lambda  \ccdot   (p_1+p_2) )\,\int d\eta \, \delta^4(p_1+p_2-\lambda \eta) 
\, \Phi(p_1,p_2)= } \\ &&
\frac{1}{(2\pi)^3} \int\!\!\!\!\int d^4p'_1\, d^4p'_2 \, 
V(p_1,p_2,p'_1,p'_2)\, \delta(p'_1\,^2
-m^2) \,  \delta(p'_2\,^2-m^2) \, (p'_1+p'_2)^2  \, 
 \int d\eta' \, \delta^4(p'_1+p'_2-\lambda \eta')\,\Phi(p'_1,p'_2).
\end{eqnarray}
It is noticed that the extra term at the meson propagator in eq.~(\ref{5j}), 
$(\lambda  \cdot   (p_2-p'_2) )^2$, is a consequence of the kinematical
character of the boost transformation in the ``point form'' formalism. 
While the mass operator corresponding to the above interaction is 
independent of the velocity, as it should be, the appearance of 
$\lambda^{\mu}$ is somewhat unusual  from a field-theory point 
of view. It leads to specific off-shell effects that change the asymptotic 
dependence from $Q^2$ to $Q^4$ in the meson propagator 
appearing in the Born amplitude, see eq.~(\ref{4m}). 
It partly explains the too fast drop-off of form factors
calculated in the impulse approximation in the ``point form'' approach.

A minimal set of two-body currents can 
be obtained by calculating the divergence of the current accounted for 
in impulse approximation,
\begin{equation}
  \frac{1}{ \sqrt{2 \, \lambda_f  \ccdot   p} } \,
\Big( 2\,\lambda_f^{\mu} \, \lambda_f  \ccdot   p + 
2\,\lambda_i^{\mu} \, \lambda_i  \ccdot   p -2\,p^{\mu} \Big) 
 \, \frac{1}{ \sqrt{2 \, \lambda_i  \ccdot  p} } \,  
\delta(\vec{q}+\vec{P_i}-\vec{P_f}), 
\label{5l}
\end{equation}
and determining what is 
needed to recover current conservation, similarly to what was done in the
non-relativistic case, see eqs.~(\ref{5c}-\ref{5e}).
Using the relation $q^{\mu}=M_f\,\lambda_f^{\mu}-M_i\,\lambda_i^{\mu}$, 
it is found that the four-divergence of the single-particle current can be 
expressed in terms of the interaction and is given by:
\begin{equation}
q_{\mu}  \ccdot   J^{\mu}_{I.A.}=  
\sqrt{\frac{m}{\lambda_f \ccdot p_{1f}}\,\frac{m}{\lambda_f \ccdot p_{2f}}}\, 
g^2 \,  \delta(\vec{q}+\vec{P_i}-\vec{P_f}) \,
\sqrt{\frac{m}{\lambda_i \ccdot p_{1i}}\,\frac{m}{\lambda_i \ccdot p_{2i}}}\,
\lambda_i \ccdot  \lambda_f\,
\left( \frac{\lambda_f  \ccdot   p_{2f}}{\lambda_i  \ccdot   p_{2f}} \, 
\frac{1}{H(\lambda_i)}
- \frac{1}{H(\lambda_f)} \, \frac{\lambda_i  \ccdot   p_{2i}}{\lambda_f  \ccdot  
 p_{2i}} \right),
\label{5m}
\end{equation}
where $H(\lambda)=\mu^2-(p_{2i}-p_{2f})^2+(\lambda \cdot (p_{2i}-p_{2f}) )^2$. 
The term of eq.~(\ref{5m}) has to be canceled by the contribution of a 
two body-current, which 
has to be guessed for some part. A possible solution is given by:
\begin{eqnarray}
\label{5n}
\nonumber 
\lefteqn{
J^{\mu}_{int}(q,p_{1i},p_{2i},p_{1f},p_{2f})= 
\sqrt{\frac{m}{\lambda_f \ccdot p_{1f}}\,\frac{m}{\lambda_f \ccdot p_{2f}}} \,
g^2 \, \delta(\vec{q}+\vec{P_i}-\vec{P_f})\,
\sqrt{\frac{m}{\lambda_i \ccdot p_{1i}}\,\frac{m}{\lambda_i \ccdot p_{2i}}}  } 
\\  &&
\qquad \times \left[ \left( 
\frac{(\lambda_f    \ccdot    p_{2f}) \, 
\lambda_f^{\mu}- p_{2f}^{\mu}   }{M_i \, (\lambda_i  \ccdot  p_{2f})} \, 
\frac{1}{H(\lambda_i)} 
+ \, \frac{1}{H(\lambda_f)} 
\frac{(\lambda_i  \ccdot   p_{2i})\,\lambda_i^{\mu}- p_{2i}^{\mu}   
}{M_f\,(\lambda_f  \ccdot   p_{2i})} \right)
+ \frac{1}{H(\lambda_f)} \,X^{\mu} \, \frac{1}{H(\lambda_i)}\right]. 
\end{eqnarray}
with
\begin{equation}
X^{\mu}=\Big( (\lambda_i + \lambda_f ) \ccdot   (p_{2i}-p_{2f}) \Big)
\,
\left(\frac{M_f\,p_{2f}^{\mu}-M_i\,p_{2i}^{\mu}}{M_f\,M_i} 
-\frac{M_f\,\lambda_f^{\mu}+M_i\,\lambda_i^{\mu}}{M_f\,M_i\,(M_i+M_f)} 
(M_f\,\lambda_f  \ccdot   p_{2f}-M_i\,\lambda_i 
 \ccdot   p_{2i}) \right).\label{5o}
\end{equation}
Terms retained here resemble those obtained by using the minimal 
coupling principle but the output, depending on the order of the 
operators, is not necessarily unique, holding up to gauge-invariant terms 
proportional 
to $\lambda_i \ccdot \lambda_f \,(M_f \,\lambda_f^{\mu}+ 
M_i\,\lambda_i^{\mu}) -(M_f\,\lambda_i^{\mu} + M_i\,\lambda_f^{\mu})$ for 
instance. This uncertainty especially affects the last term in  
eq.~(\ref{5o}). It can be removed for a part by requiring 
to recover the Born amplitude (see below). 

The above current could solve some of the problems at low $Q^2$ 
related to current conservation. Again, we notice that it
does not contain any $1/Q^2$ factor. The appearance of the total mass 
$M$ at the denominator in the above two-body current is not quite 
expected and, most probably, results from enforcing current conservation. 
By itself, the presence of this quantity in the current is not surprising, 
as it is well known that the momentum in the ``point form'' approach, 
and therefore the current, depends on the interaction. However, it can 
be checked that the above current does not help in solving the vanishing 
of the elastic scalar form factor in the limit $M\rightarrow 0$, which 
also shows up at small momentum transfers. 

\subsubsection{Two-body currents motivated by the  ratio $F_0/F_1$ 
in the limit $M\rightarrow 0$}

\noindent
While considering the problem, we found that the required 
two-body currents also contribute to the norm, defined by 
the charge associated with the conserved current, $F_1(q^2=0)=1$. This 
differs from standard two-body currents, in the two-nucleon system for 
instance, or in eq.~(\ref{5e}), which do not contribute to the charge (in the
usual approach where the interaction does not depend on the energy). 
Extra currents 
involve double Z-diagrams, like the one shown in Fig.~\ref{fig3}c. They are not 
especially suppressed for the scalar coupling model considered here 
and they  contribute destructively to the norm so that to cancel 
the normal contribution (Fig.~\ref{fig3}a) in the limit $M\rightarrow 0$. 
Such a result can be checked by calculating the contribution 
of the triangle diagram of the figure in the case where the momentum 
dependence of the bound state vertex function is ignored (see also 
ref.~\cite{Coester:1994wp}):
\begin{equation}
\int d^4p \,\frac{P_i^0+P_f^0-2\,p^0}{m^2-p^2-i\epsilon}
\frac{1}{m^2-(P_i-p)^2-i\epsilon} \frac{1}{m^2-(P_f-p)^2-i\epsilon} 
=2i\pi \int d^3p \, \frac{1}{(2\,e_p)^3}\, 
\left( \frac{2\,e_p}{(2\,e_p-M)^2} - \frac{2\,e_p}{(2\,e_p+M)^2}\right),
\label{5p}
\end{equation}
where $P_i^0$ and $P_f^0$ are expressed in the c.m. system, $P_i^0=P_f^0=M$.
For the scalar case, one gets:
\begin{eqnarray}
\label{5q}
\nonumber
\lefteqn{
\int d^4p \, \frac{2\,m}{m^2-p^2-i\epsilon} \, \frac{1}{m^2-(P_i-p)^2-i\epsilon}
\, \frac{1}{m^2-(P_f-p)^2-i\epsilon} =} \\ && \qquad\qquad 2i\pi \int d^3p 
 \frac{1}{(2\,e_p)^3}\, 
\left( \frac{2\,m}{(2\,e_p-M)^2} + \frac{4\,m}{2\,e_p\,(2\,e_p-M)} + 
\frac{4\,m}{2\,e_p\,(2\,e_p\,+M)} + \frac{2\,m}{(2\,e_p+M)^2} 
\right).
\end{eqnarray}
\end{widetext}
In eqs.~(\ref{5p},\ref{5q}), the first term at the r.h.s. represents the 
standard non-relativistic contribution. The extra terms, which should 
also occur in the instant form of relativistic quantum mechanics, greatly 
complicate the calculation of form factors. They cannot be neglected however and, 
in fact, their introduction seems to provide more consistency in the 
developments. The matrix element of the current in the ``point form'' approach, 
eq.~(\ref{2q}), appears to be 
proportional to the factor $M$ because this one has been introduced as an 
overall factor. In the Bethe-Salpeter approach, eq. (3), this factor appears 
dynamically. The fact that the four-vector current matrix element should be 
proportional to $P_i^{\mu}+P_f^{\mu}$ automatically ensures that it is 
proportional to  $M$ at $\vec{P}=0$ (without requiring the introduction 
of this factor by hand). This is just a consequence of the extra 
current discussed above. 

At the same time as the front factor $M$ is removed from the r.h.s. 
of eq.~(\ref{2q}), it disappears from the expression of the scalar 
form factor, eq.~(\ref{2p})  and eq.~(\ref{2s}). This form factor 
does not vanish anymore in the limit $M\rightarrow 0$. The ratio 
$F_0(0)/F_1(0)$ stemming from eqs.~(\ref{5p},\ref{5q}) is $1.5$ in the limit 
$M\rightarrow 0$, while the Wick-Cutkosky result is $1.25$\footnote{Notice that the 
light-front approach 
seems to do correctly with respect to this problem. The ratio
$F_0(0) / F_1(0)$ is finite in the limit $M\rightarrow 0$ and 
its value, $7/6$, is close to the expected one. Furthermore,
a contribution like the double Z-diagram of Fig.~\protect\ref{fig3}c vanishes 
in this approach for the model considered here.}.  
Finally, to recover 
the full Born amplitude, one can use the standard definition of the 
current without renormalizing its expression, somewhat arbitrarily, 
by a factor $1/M$ in order to compensate the front factor $M$ 
in eqs.~(\ref{2p},\ref{2s}).

The discussion of the above contributions would require a full paper by itself 
and, as far as we can see, they do not help in solving the current conservation 
problem considered in this subsection. In practice, we will account for them by 
multiplying the single-particle current operator by a constant factor suggested 
by the expression of eq.~(\ref{5p}):
\begin{equation}
F= 1- \left(\frac{M-2\,\bar{e}}{M+2\,\bar{e}}\right)^2
= \frac{8\,M\,\bar{e}}{(M+2\,\bar{e})^2},
\label{5r}
\end{equation}
where $\bar{e}$ represents an average value of $\sqrt{m^2+p^2}$. 
The two-body nature of the correction is not explicit, but corresponding to an 
off-shell effect, it can be made transparent by expressing the factor 
$M-2e$ in terms of the interaction using eq.~(\ref{5b}). The correction is 
at least of the second order in this interaction and while its effect is small 
for weakly bound systems, it it certainly considerable when the total mass 
goes to 
zero.  Interestingly, the above example gives a further illustration of how a 
term proportional to the kinetic energy, $e_p$ (first term in the r.h.s. of 
eq.~(\ref{5p})) turns into a term proportional to the total mass by 
incorporating interaction effects.

\begin{figure}[tb]
\begin{center}
\includegraphics[width=\columnwidth]{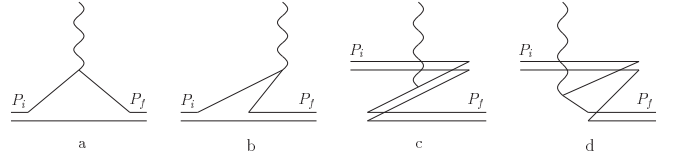}
\end{center}
\caption{Time-ordered triangle diagrams contributing to the absorption of a 
photon on  a bound system.\label{fig3}}
\end{figure}

The above approximation would certainly be questionable for an exact 
calculation but, being interested in whether some of the striking 
features evidenced by the impulse ``point form'' results can be 
repaired for some part by adding two-body currents, we believe 
it should not affect the developments presented below. On the other hand, 
it allows one to continue to work with a conserved current, 
including the single- and the two-body parts 
given by eqs.~(\ref{5n},\ref{5o}). 

%%%%%%%%%%%%%%%%%%%%
\subsubsection{Two-body currents motivated by the  Born amplitude in the 
``point form'' approach}

\noindent
When considering the further requirement of reproducing the Born amplitude, an 
extra contribution arises. This one is obtained by subtracting from this  
amplitude the contribution accounted for in the impulse approximation 
calculation, eqs.~(\ref{2p},\ref{2q}), and that one accounting for current 
conservation, eqs.~(\ref{5n},\ref{5o}) (see also appendix~\ref{app:e}). 
As current conservation holds for the Born amplitude, it also holds 
for the above difference in the same limit, i.e. at the order $g^2$. Neglected 
contributions are of the order $g^4$, which in any case are discarded when 
limiting ourselves to the Born amplitude. A few details are given in  
appendix~\ref{app:e}, here we give an expression where the two-body 
contribution is written in a way where current conservation is manifest, 
analogously to eq.~(\ref{5h}): 
\begin{eqnarray}
\label{5s}
\nonumber
\lefteqn{
\epsilon_{\mu} \ccdot J^{\mu}_{\Delta B}(q,p_{1i},p_{2i},p_{1f},p_{2f})= 
} \\ \nonumber && 
\sqrt{\frac{m}{\lambda_f \ccdot p_{1f}}\,\frac{m}{\lambda_f \ccdot p_{2f}}}  
\frac{g^2}{4}\delta(\vec{q}+\vec{P_i}-\vec{P_f}) 
\sqrt{\frac{m}{\lambda_i \ccdot p_{1i}}\,\frac{m}{\lambda_i \ccdot p_{2i}}}
\\ \nonumber && \times 
\frac{\mu^2-(p_{2i}-p_{2f})^2+\lambda_i \ccdot (p_{2i}-p_{2f}) 
\, \lambda_f \ccdot (p_{2f}-p_{2i})  }{\lambda_f \ccdot p_{2f} \, \lambda_i 
\ccdot p_{2i} \, H(0)  \, H(\lambda_f)  H(\lambda_i)} 
\\  && \times \, \frac{2\,\bar{e}}{M}  \epsilon_{\mu} \ccdot 
\Big( Y^{\mu}     (p_{2i}-p_{2f})\ccdot  q
  -     (p_{2i}-p_{2f})^{\mu}\, Y \ccdot  q \Big), 
\end{eqnarray}
with
\begin{equation}
Y^{\mu}=\lambda_f^{\mu} \, \lambda_f \ccdot p_{2f}
    + \lambda_i^{\mu} \, \lambda_i \ccdot p_{2i}
    -\frac{1}{2}(p_{2f}^{\mu}  +p_{2i}^{\mu}).\nonumber
\label{5t}
\end{equation}
In deriving this expression, we assume the relationship:
\begin{equation}
q^{\mu}=\lambda_f^{\mu}\,M_f- \lambda_i^{\mu}\,M_i   \simeq
\frac{M}{\bar{e}}\,(\lambda_f^{\mu}\,\lambda_f \ccdot p_{2f}  
       - \lambda_i^{\mu} \, \lambda_i \ccdot p_{2i}),
\label{5u}
\end{equation}
where the replacement of $\lambda \ccdot p$ by an average value $\bar{e}$ is in 
accordance with neglecting contributions of order higher in $g^2$. 
    
For the scalar probe, the extra contribution required to reproduce the Born 
amplitude is given by:
\begin{eqnarray}
\label{5v}
\nonumber
\lefteqn{
S_{\Delta B}(q,p_{1i},p_{2i},p_{1f},p_{2f})= } 
\\ && \nonumber
\sqrt{\frac{m}{\lambda_f \ccdot p_{1f}}\,\frac{m}{\lambda_f \ccdot p_{2f}}} \, 
\frac{g^2}{2}\,\delta(\vec{q}+\vec{P_i}-\vec{P_f}) \, 
\sqrt{\frac{m}{\lambda_i \ccdot p_{1i}}\,\frac{m}{\lambda_i \ccdot p_{2i}}}
\\ && \times \, \left[\,
 \frac{1}{H(0)} \,
\left( \frac{m}{\lambda_i  \ccdot   p_{2i}\,\lambda_i  \ccdot   p_{2f}\,} 
+ \frac{m}{\lambda_f  \ccdot   p_{2f}\,\lambda_f  \ccdot   p_{2i}\,} 
\right) \right. \nonumber \\ &&
\left. - 
\left( \frac{m \, \lambda_i  \ccdot   (p_{2i}-p_{2f}) }{\lambda_i  \ccdot   
p_{2f}\, H(0) \,H(\lambda_i)}
 + \frac{m \, \lambda_f  \ccdot   (p_{2f}- p_{2i}) }{\lambda_f  \ccdot   
p_{2i}\, H(0) 
\,H(\lambda_f)}  \right) \, \right]. \nonumber \\
\end{eqnarray}
Notice that eqs.~(\ref{5s},\ref{5v}) correspond to 
contributions to the Born amplitude that are not generated in another 
way and that there is therefore no double counting.  Examination 
of the two-body currents, eqs.~(\ref{5n},\ref{5o}) 
and~(\ref{5s},\ref{5v}), which involve a neutral boson exchange,
shows that they are much more sophisticated than standard ones in 
the same case. They exhibit unusual features, like the appearance 
of the boson propagator twice, which generally characterizes 
the contribution of a charged boson interacting with an external field.

%%%%%%%%%%%%%%%%%%%%%%%%%%%%%%5.2
\subsection{Results involving two-body currents}
\label{subsec:52}

\noindent
In this subsection, we present expressions for form factors 
incorporating contributions from two-body currents. These ones are 
motivated by fulfilling current conservation and reproducing 
the Born amplitude. They are followed by two sets of results, 
obtained with a Galilean boost and the ``point form'' one. In both cases, 
the wave function from the model v1 is used.

%%%%%%%%%%%%%%%%%%%%%%
\subsubsection{Expressions of the calculated form factors}

\noindent
The two body-currents employed with the Galilean boost are inspired 
from eqs.~(\ref{5e},\ref{5h}) for the parts required to fulfill 
current conservation and reproduce the Born amplitude, respectively. 
However, since these currents are appropriate to an instant-form formalism, 
we use an alternative expression of these currents that we could derive
assuming a Galilean invariant extension of the interaction model v1. 
These ones miss the relation to a well defined field-theory motivated 
current and evidence features that are sometimes unusual with this respect, 
although a relation to pair-type currents can be recovered in some limit. 
As they are not of fundamental importance and perhaps too specialized, 
we prefer to give their expressions together with the contributions 
to form factors in  appendix~\ref{app:d}. 

For the ``point form'' approach, results are obtained from 
eqs.~(\ref{5n},\ref{5o},\ref{5r},\ref{5s},\ref{5v}). 
The full expressions of the different form factors, including 
two-body currents, are given below, while some intermediate steps 
are given in appendix~\ref{app:e}. Using the relation 
implied by the $\delta^4$ functions, the momenta relative to particle 
number 1  can always be written in terms of the momenta relative to particle 
number 2 (the spectator particle) and the four-velocity vector $\lambda^{\mu}$. 
In absence of ambiguity, the momentum
$\vec{p}$ in the expressions given below will refer to this spectator 
particle. When specialized to the frame $\vec{v}=\vec{v}_f=-\vec{v}_i$, 
the form factors $F_0(q^2)$, $\tilde{F}_1(q^2)$ and 
$\tilde{F}_2(q^2)$ successively read:
\begin{widetext}
\begin{eqnarray}
\nonumber
\lefteqn{
F_0(q^2)= \frac{\sqrt{N_f\,N_i}}{4\,m} 
\left(\int \frac{d \vec{p}}{(2\pi)^3} 
\,\phi_f(\vec{p}_{tf}) \, \frac{m}{e_p} \, \phi_i(\vec{p}_{ti}) 
+ \int \frac{ d\vec{p}_f\,d\vec{p}_i}{(2\pi)^6}
\,  \phi_f(\vec{p}_{tf})\,\phi_i(\vec{p}_{ti}) \, 
\frac{m^2}{e_{f}\,e_{i}}\, (K_{\Delta B})_0 \right),} \\
\nonumber
\lefteqn{
\tilde{F}_1(q^2)= \frac{\sqrt{N_f\,N_i}}{2\,\bar{M}}\, 
\sqrt{1-v^2} \left[   \left( 1- (\frac{\bar{M}-2\,\bar{e}}{\bar{M}+2\,\bar{e}})^2 \right)
\frac{1+v^2}{1-v^2} \int \frac{d \vec{p}}{(2\pi)^3} 
\,\phi_f(\vec{p}_{tf}) \,\phi_i(\vec{p}_{ti})
\right. } \\ &&  \left.
+    \left( 1- (\frac{\bar{M}-2\,\bar{e}}{\bar{M}+2\,\bar{e}})^2 \right)
\int \frac{ d\vec{p}_f\,d\vec{p}_i}{(2\pi)^6}
\,  \phi_f(\vec{p}_{tf})\,\phi_i(\vec{p}_{ti}) \, 
\frac{m^2}{e_{f}\,e_{i}}\, (K_{int})_1
+ \int \frac{ d\vec{p}_f\,d\vec{p}_i}{(2\pi)^6}
\,  \phi_f(\vec{p}_{tf})\,\phi_i(\vec{p}_{ti}) \, 
\frac{m^2}{e_{f}\,e_{i}}\, (K_{\Delta B})_1  \right], \nonumber \\
\nonumber
\lefteqn{
\tilde{F}_2(q^2) \, \vec{v}= -\frac{\sqrt{N_f\,N_i}}{2\,\bar{M}}\, 
\sqrt{1-v^2}\left[  
\left( 1- (\frac{\bar{M}-2\,\bar{e}}{\bar{M}+2\,\bar{e}})^2 \right) \, 
\frac{1+v^2}{1-v^2} \int \frac{d \vec{p}}{(2\pi)^3} 
\, \phi_f(\vec{p}_{tf}) \,\,  \frac{\vec{p}}{e_p} \,\, 
\phi_i(\vec{p}_{ti})\right. } \\ && \left.
+ \left( 1- (\frac{\bar{M}-2\,\bar{e}}{\bar{M}+2\,\bar{e}})^2 \right) \, 
\int \frac{ d\vec{p}_f\,d\vec{p}_i}{(2\pi)^6}
\,  \phi_f(\vec{p}_{tf})\,\phi_i(\vec{p}_{ti}) \, 
\frac{m^2}{e_{f}\,e_{i}}\, (\vec{K}_{int})_2 
+ \int \frac{ d\vec{p}_f\,d\vec{p}_i}{(2\pi)^6}
\,  \phi_f(\vec{p}_{tf})\,\phi_i(\vec{p}_{ti}) \, 
\frac{m^2}{e_{f}\,e_{i}}\, (\vec{K}_{\Delta B})_2  \right],
\label{5w}
\end{eqnarray}
where $\bar{M}$, $\bar{e}$ and $N$ are defined in appendix~\ref{app:d}.
The expressions of the $K$ quantities, which account for two-body currents, are
given by:
\begin{eqnarray}
(K_{int})_1&=& g^2  \,  
\left[ \left( 
\frac{(\lambda_f    \ccdot    p_{f}) \, 
\lambda_f^{0}- p_{f}^{0} }{M_i \,(\lambda_i  \ccdot  p_{f}) \, H(\lambda_i)}
+ \frac{(\lambda_i  \ccdot   p_{i})\,\lambda_i^{0}- p_{i}^{0}   
}{M_f \, H(\lambda_f) \, (\lambda_f  \ccdot   p_{i})} \right)
+ \frac{1}{H(\lambda_f)} \,X^0 \, \frac{1}{H(\lambda_i)}\right], 
\nonumber \\
(\vec{K}_{int})_2&=& g^2 \,
\left[ \left( \frac{(\lambda_f    \ccdot    p_f) \, \vec{\lambda}_f- \vec{p}_f   
}{M_i \, (\lambda_i  \ccdot  p_f) \, H(\lambda_i)}
+ \frac{(\lambda_i  \ccdot   p_i)\,\vec{\lambda}_i- \vec{p}_i  
}{M_f \, H(\lambda_f)\,(\lambda_f  \ccdot   p_i)} \,  \right)
+ \frac{1}{H(\lambda_f)} \,\vec{X} \, \frac{1}{H(\lambda_i)}\right], 
\nonumber \\
(K_{\Delta B})_0&=&\frac{g^2}{2}
  \, \left[\,
 \frac{1}{H(0)} \,
\left( \frac{m}{\lambda_i  \ccdot   p_{i}\,\lambda_i  \ccdot   p_{f}\,} 
+ \frac{m}{\lambda_f  \ccdot   p_{f}\,\lambda_f  \ccdot   p_{i}\,} 
\right)  - 
\left( \frac{m \, \lambda_i  \ccdot   (p_{i}-p_{f}) }{\lambda_i  \ccdot   
p_{f}\, H(0) \,H(\lambda_i)}
 + \frac{m \, \lambda_f  \ccdot   (p_{f}- p_{i}) }{\lambda_f  \ccdot   
p_{i}\, H(0) \,H(\lambda_f)}  \right) \, \right], 
\nonumber \\
(K_{\Delta B})_1&=& \frac{g^2}{4}\,  
\frac{\mu^2-(p_i-p_f)^2+\lambda_i \ccdot (p_i-p_f) 
\, \lambda_f \ccdot (p_f-p_i)  }{\lambda_f \ccdot p_f \, \lambda_i 
\ccdot p_i \, H(0)  \, H(\lambda_f) \, H(\lambda_i)} 
  \, \frac{2\,\bar{e}}{\bar{M}} \,
\left( Y^{0} \,   (p_i-p_f)\ccdot q
  -     (p_i-p_f)^{0}\, Y \ccdot  q\right),  \nonumber 
\\
(\vec{K}_{\Delta B})_2&=& \frac{g^2}{4}\,  
\frac{\mu^2-(p_i-p_f)^2+\lambda_i \ccdot (p_i-p_f) 
\, \lambda_f \ccdot (p_f-p_i)  }{\lambda_f \ccdot p_f \, \lambda_i 
\ccdot p_i \, H(0)  \, H(\lambda_f) \, H(\lambda_i)} 
 \, \frac{2\,\bar{e}}{\bar{M}} \,
\left( \vec{Y} \,   (p_i-p_f) \ccdot q
  -     (\vec{p}_i-\vec{p}_f)\, Y \ccdot  q\right). 
\label{5x}
\end{eqnarray}
\end{widetext}
As it can be observed, the expression of the currents required 
to ensure current conservation ($K_{int},\vec{K}_{int}$) do not 
contain any $1/q^2$ factor as the recipe given by eq.~(\ref{5g}) 
would imply.

%%%%%%%%%%%%%%%%%%%%%%%
\subsubsection{Results  with two-body currents for the  form factors 
in a Galilean approach}

\noindent
The non-relativistic type calculations of table~\ref{t50} are especially 
useful to make the transition from the ``exact'' results presented 
in sect.~\ref{sect:3} to the ``point form'' ones, allowing one to distinguish effects 
specific of this last approach from those due to the restoration of current 
conservation, to the Born amplitude constraint or to the dynamics. 

\begin{table}[tb]
\caption{Elastic and inelastic form factors $F_0(q^2)$, 
$F_1(q^2)$ and  $F_2(q^2)$ in a Galilean approach as given in 
appendix~\protect\ref{app:d}: Effect 
of two-body currents motivated  by current conservation ($int$ line) 
and the Born amplitude ($int+\Delta B$ line). Calculations are 
performed with the interaction model v1 and correspond to a coupling 
$\alpha=3$ for the Wick-Cutkosky model.}
\label{t50}
\begin{ruledtabular}
\begin{tabular}{lccccc}
  $Q^2/m^2$     &   0.01   &   0.1  &  1.0   &  10.0 & 100.0 \\ [1.ex] \hline
 \multicolumn{3}{l}{$\alpha=3, \,elastic,\,int $}  \\ [0.ex] 
% $\alpha=3, \,elastic,\,int $   &   &   &        &       &  \\ [0.ex] 
 $F_0$      & 0.735    & 0.701  & 0.454  &  0.365-01  & 0.198-03  \\ [0.ex] 
 $F_1$      & 0.995    & 0.953  & 0.640  &  0.665-01  & 0.774-03  \\ [1.ex] 
\multicolumn{3}{l}{$\alpha=3, \,elastic,\,int+\Delta B $}  \\ [0.ex] 
% $\alpha=3, \,elastic,\,int+\Delta B $   &   &   &        &       &  \\ [0.ex] 
 $F_0$      & 0.959    & 0.921  & 0.628  &  0.635-01  & 0.643-03  \\ [0.ex] 
 $F_1$      & 0.996    & 0.953  & 0.641  &  0.680-01  & 0.898-03  \\ [1.ex] 
\multicolumn{3}{l}{$\alpha=3, inelastic,\,int$}  \\ [0.ex]
% $\alpha=3, inelastic,\,int$   &   &   &        &       &  \\ [0.ex] 
 $F_0$   & -0.325-01 & 0.202-02 & 0.132-00 &  0.189-01  & 0.91-04  \\  [0.ex]    
 $F_1$   & 0.469-02  & 0.430-01 & 0.186-00 &  0.318-01  & 0.33-03  \\  [0.ex]    
 $F_2$   & 0.483     & 0.442    & 0.191-00 &  0.327-02  & 0.34-05  \\  [0.ex] 
 \multicolumn{3}{l}{(two-body part of $F_2$)} \\
   & (24\%) & (25\%) & (28\%)  &  (48\%)  & (116\%)  \\  
[1.ex] 
 \multicolumn{3}{l}{$\alpha=3, inelastic,\,int+\Delta B$}  \\ [0.ex]   
% $\alpha=3, inelastic,\,int+\Delta B$   &   &   &        &       &  \\ [0.ex] 
 $F_0$   & 0.086-01 & 0.509-01  & 0.209-00  &  0.361-01  & 0.37-03  \\  [0.ex]    
 $F_1$   & 0.468-02 & 0.429-01  & 0.185-00  &  0.325-01  & 0.41-03  \\  [0.ex]    
 $F_2$   & 0.481    & 0.442     & 0.191-00  &  0.335-02  & 0.42-05  \\  [1.ex] 
\end{tabular}
\end{ruledtabular}
\end{table}

At low momentum transfers, the suppression of the elastic form factor 
$F_0$ with respect to $F_1$ ($int$ case) is reminiscent for some part 
of the one mentioned in sect.~\ref{sect:3} for the ``point form'' case as a 
result of the normalization definition. It is largely canceled by a 
pair term contribution to the scalar form factor included in the 
Born-constrained current. The same contribution also changes the 
scalar inelastic form factor $F_0$ from a negative to a positive value. 
In both cases, the results (0.959 and 0.009 at   $Q^2/m^2=0.01$) 
become closer to the ``exact'' ones (1.123 and 0.054). 
At high momentum transfers, the elastic and inelastic form factors, 
$F_0$, in the $int$ case decrease more quickly than the corresponding 
vector  form factors, $F_1$. This is due to an extra $1/Q$ dependence 
in the former. This discrepancy tends to disappear when the 
Born-constrained current is considered. The results so obtained 
are qualitatively in better agreement with the ``exact'' ones. 
However the magnitude is smaller by a factor 5 or so. The comparison 
with the Coulombian-type results suggests that the discrepancy is 
to be ascribed to a large part to the difference in the wave functions 
at the origin that, as already mentioned, determines the overall 
coefficient multiplying the asymptotic power law behavior 
of form factors. This points to the relative simplicity of the 
interaction model, v1, which does not do quite well as to 
the description of the spectrum of the Wick-Cutkosky model, 
while the Coulombian one does. We do not expect this discrepancy 
to be removed when looking at the ``point form'' results. 

The last comment we want to make concerns the current conservation 
that is better seen by looking at the form factor, $F_2$. 
Contrary to the Born-constrained current, the two-body current motivated by 
current conservation has a big influence in the
$int$ case. From 25\% of $F_2$ at low momentum transfers, 
its contribution can raise up to 50\% for higher momentum transfers. 
Larger contributions are expected when the average momentum 
of the constituent particles increases ($\bar{p}^2/m^2=0.2$ in the 
present case).

%%%%%%%%%%%%%%%%%%%%%%%%
\subsubsection{Results with two-body currents for the  form factors 
in the ``point form'' approach}

\noindent
The least that one can say about the contributions of two-body currents 
in the ``point form'' approach is that they strongly depend on the coupling, 
the mass of the system, the elastic or inelastic,  scalar or vector character 
of the transition, and the presence or absence of a node in the impulse 
approximation. We successively consider the effect of two-body currents 
motivated by current conservation and the Born amplitude behavior, 
the corresponding results being given in table 
\ref{t60}\footnote{Involving a 4-dimensional 
integration, the two-body part of the form factors involves some uncertainty 
and the accuracy of the results may be smaller than what the number 
of digits suggests. Differences may be significant however.}. 

Beginning with the elastic form factor corresponding to $\alpha=3$, 
we found that the effect of two-body 
currents motivated by current conservation is quite small at low $Q^2$ 
but, being constrained to go to 0 at zero momentum transfer with $Q^2$, 
this result is not very significant. At higher $Q^2$, the effect is 
much larger (40\% of the impulse approximation at $Q^2=10\,m^2$, more 
than 100\% at $Q^2=100\,m^2$). In comparison to the results given 
in table~\ref{t50} in a similar case, we nevertheless notice that 
the drop-off of the form factor is faster, scaling like $Q^{-6}$ 
rather than $Q^{-4}$ at high $Q^2$. 
The inelastic form factor is more instructive 
on the effects at low $Q^2$ where  $F_2(Q^2)$ is enhanced by 
about 30\% while $F_1(Q^2)$ gets decreased by 100\% or so. 
The large effects in this case are strongly related to current 
conservation which implies that $F_1(Q^2)$ scales like $Q^2$ at small 
momentum transfer. At higher momentum transfer, 100\% effects, 
constructive or destructive, are also found. The last case now concerns 
the $F_2(Q^2)$ form factor which is seen to scale like $Q^{-8}$ while 
$F_1(Q^2)$ scales like $Q^{-6}$. In comparison to the Galilean calculation 
presented in table~\ref{t50}, the relative size of the corrections 
is qualitatively the same. As for the elastic case, the form factor
$F_1(Q^2)$ drops too fast ($Q^{-6}$ instead of $Q^{-4}$). The results 
for the strongly interacting case ($M=0.1$) provide a useful and 
complementary information.  The corrections make the form factor  
overshoot the value $F_1(Q^2=0)=1$. Essentially, depending on the quantity 
$v^2$, the corrections scale like $M^{-2}$. They are therefore enhanced 
in the small mass limit, in the same way that the charge radius is 
in the present ``point form'' approach. The correction drops off quickly 
at higher $Q^2$ but this shows the limitations that underlie the 
derivation of two-body currents to which too much is asked in the 
present case. 

\begin{table}[tb]
\caption{Elastic and inelastic form factors $F_0(q^2)$, $F_1(q^2)$ and
$F_2(q^2)$, calculated in the ``point form'' approach and including two-body 
currents: Results of eqs.~(\protect\ref{5w},\protect\ref{5x}) 
are calculated with the interaction model, v1, and are 
given for different couplings of the Wick-Cutkosky 
model, $\alpha=3$ ($\alpha(v1)=1.775$, $E= 0.432\,m$) 
and $\alpha\simeq 2\,\pi$ ($\alpha(v1)=5.287$, $E=1.90\,m$). 
Different approximations about the two-body currents are considered, 
successively: impulse approximation ($I.A.$), with inclusion of 
interaction currents ($int$) and with inclusion of both interaction and 
Born motivated currents ($int+\Delta B$).}
\label{t60}
\begin{ruledtabular}
\begin{tabular}{lccccc}
  $Q^2/m^2$     &   0.01   &   0.1  &  1.0   &  10.0 & 100.0 \\ [1.ex] \hline
\multicolumn{3}{l}{$\alpha=3, \,elastic,\, I.A.$} \\ [0.ex] 
% $\alpha=3, \,elastic,\, I.A.$   &   &   &        &       &  \\ [0.ex] 
 $F_0$      & 0.732    & 0.671  & 0.312  &  0.061-01  & 0.29-05  \\ [0.ex] 
 $F_1$      & 0.992    & 0.924  & 0.493  &  0.205-01  & 0.46-04  \\ [1.ex] 
\multicolumn{3}{l}{$\alpha=3, \,elastic,\,int$} \\ [0.ex] 
% $\alpha=3, \,elastic,\,int$   &   &   &        &       &  \\ [0.ex] 
 $F_0$      & 0.732    & 0.671  & 0.312  &  0.061-01  & 0.29-05  \\ [0.ex] 
 $F_1$      &  0.992   & 0.930  & 0.521  &  0.295-01  & 0.11-03  \\ [1.ex] 
\multicolumn{3}{l}{$\alpha=3, \,elastic,\,int+\Delta B$} \\ [0.ex] 
% $\alpha=3, \,elastic,\,int+\Delta B$   &   &   &        &       &  \\ [0.ex] 
 $F_0$      & 1.142    & 1.065  & 0.573  &  0.268-01  & 0.20-03  \\ [0.ex] 
 $F_1$      & 0.994   & 0.944 & 0.595  &  0.491-01  & 0.43-03  \\ [1.ex] 
\multicolumn{3}{l}{$\alpha=3, inelastic,\,I.A.$} \\ [0.ex] 
% $\alpha=3, inelastic,\,I.A.$   &   &   &        &       &  \\ [0.ex] 
 $F_0$     & 0.011    & 0.043  & 0.116  &  0.489-02  & 0.26-05  \\  [0.ex]    
 $F_1$     & 0.019    & 0.059  & 0.167  &  0.136-01  & 0.33-04  \\  [0.ex]    
 $F_2$     & 0.383    & 0.335  & 0.102  &  -0.373-03  & -0.34-05  \\  [1.ex] 
\multicolumn{3}{l}{$\alpha=3, inelastic,\,int$} \\ [0.ex]   
% $\alpha=3, inelastic,\,int$   &   &   &        &       &  \\ [0.ex] 
 $F_0$     & 0.011    & 0.043  & 0.116  &  0.489-02  & 0.26-05  \\  [0.ex]    
 $F_1$     & 0.006  & 0.047  & 0.168  &  0.181-01  & 0.73-04  \\  [0.ex]    
 $F_2$     & 0.535    & 0.474  & 0.171  &  0.186-02  & 0.79-06  \\  [1.ex] 
\multicolumn{3}{l}{$\alpha=3, inelastic,\,int+\Delta B$} \\ [0.ex] 
% $\alpha=3, inelastic,\,int+\Delta B$   &   &   &        &       &  \\ [0.ex] 
 $F_0$     & 0.059    & 0.097   & 0.187   &  0.167-01   & 0.12-03  \\  [0.ex]    
 $F_1$     & 0.005  & 0.043 & 0.170 &  0.274-01  & 0.24-03  \\  [0.ex]    
 $F_2$     & 0.469  & 0.432 & 0.173 &  0.282-02  & 0.25-05  \\  [1.ex] 
\multicolumn{3}{l}{$\alpha\simeq 2\,\pi, elastic,\,I.A.$} \\ [0.ex] 
% $\alpha\simeq 2\,\pi, elastic,\,I.A.$   &   &   &        &       &  \\ [0.ex] 
 $F_0$   & 0.096   & 0.012-01  & 0.50-06  &  0.86-10  & 0.12-13  \\  [0.ex]    
 $F_1$   & 0.486   & 0.167-01  & 0.34-04  &  0.37-07  & 0.37-10  \\  [1.ex]    
\multicolumn{3}{l}{$\alpha\simeq 2\,\pi, elastic,\,int$} \\ [0.ex] 
% $\alpha\simeq 2\,\pi, elastic,\,int$   &   &   &        &       &  \\ [0.ex] 
 $F_0$   & 0.096   & 0.001  & 0.50-06  &  0.86-10  & 0.12-13  \\  [0.ex]    
 $F_1$   & 3.831   & 0.460  & 0.20-02  &  0.23-05  & 0.24-08  \\  [1.ex]    
\multicolumn{3}{l}{$\alpha\simeq 2\,\pi, elastic,\,int+\Delta B$} \\ [0.ex] 
% $\alpha\simeq 2\,\pi, elastic,\,int+\Delta B$   &   &   &   &    &  \\ [0.ex] 
 $F_0$   & 0.359   & 0.012  & 0.92-04  &  0.87-06  & 0.87-08  \\  [0.ex]    
 $F_1$   & 5.452   & 0.699  & 0.49-02  &  0.32-04  & 0.30-06  \\  [1.ex]    
\end{tabular}
\end{ruledtabular}
\end{table}

All the above results drop off too fast in comparison to the expected 
power law behavior. This one therefore relies on the two-body currents  
added to reproduce the Born amplitude. At high  $Q^2$, the results 
so obtained are however below the ``exact'' ones given in 
tables~\ref{t10}-\ref{t30}. 
Part of the effect is certainly due to the choice of the interaction v1. 
The comparison with results given in table~\ref{t50}, 
obtained with the same interaction model, should be more  
adequate but it still shows some discrepancy. The results for the strongly 
interacting case are again useful here. The suppression of the form factor 
at high $Q^2$ by many orders of magnitude points to a dependence of the 
asymptotic form factor on a factor $M^4$, which simply stems from the 
dependence of the form factor on the velocity $v$, which involves the factor 
$Q/M$. The reduction factor $(M/(2\bar{e}))^4$ largely explains the 
difference in the results given in tables~\ref{t50} and~\ref{t60} 
for $\alpha=3$. As reminded in the beginning of sect.~\ref{sect:4}, one expects 
such factors to be canceled by interaction effects which, thus, do not 
appear to be accounted for by the two-body currents we looked at. 
The currents discussed here also contribute at low $Q^2$. 
The most significant effect concerns the scalar form factor. The 
contributions, respectively 0.4 and 0.05 for the elastic and inelastic 
form factors for $\alpha=3$ compare with those obtained in the Galilean 
calculation, 0.22 and 0.04 (see table~\ref{t50} ).

While looking at two-body currents motivated by current conservation, 
an interesting question was whether they could allow one to get the 
low $Q^2$ charge form factor right, in relation with the charge radius. 
Results for the strong interaction  case leave this possibility open  
but those for the smaller coupling, $\alpha=3$, rather point to a different 
answer. In this case, which is under better control (not much extra 
currents needed), most of the correction at low $Q^2$ is produced 
by the term introduced to ensure the right power law behavior. 

%\clearpage
%%%%%%%%%%%%%%%%%%%%%%%%%%%%%%%%66666666666%%%%%%%%%%%%%%%%%%%%%%%%%%%%%%%%%%

\section{Conclusion}
\label{sect:6}

\noindent
In this work, we further investigated the calculation of form factors 
in the ``point form'' approach for a two-body system. With respect 
to a previous work~\cite{Desplanques:2001zw}, we also considered  a scalar form factor, 
which gives another information that can be compared to a more 
elaborate calculation. This one is provided by the Wick-Cutkosky model 
where form factors can be calculated exactly as far as one neglects 
mass and vertex renormalization as usually done when dealing with a 
two-body system. Possible corrections, that could be accounted for by 
introducing form factors for the constituents,  cancel out in this
case but should be incorporated when a comparison to experiment 
is done. The consideration of the scalar form factor confirms
conclusions reached previously. Taking into account that the 
non-relativistic calculation does rather well (which by itself 
deserves some explanation), the implementation of the ``point form'' 
approach in the impulse approximation, as used in recent works,  
does badly both at low and large $Q^2$. At low $Q^2$, results are 
not protected by the conservation of some charge as for the vector case. 
At large $Q^2$ it decreases more quickly, $1/Q^8$ instead of $1/Q^4$  
(up to log factors). We also checked that, qualitatively, the results 
for both the scalar and vector form factors were unchanged by using 
another interaction in the mass operator. The limit $M \rightarrow 0$ 
reveals that the charge and scalar radii scale like $1/M$ 
in the ``point form'' approach while the scalar form factor tends to $0$ 
when $Q^2 \rightarrow 0$. None of these features is supported by 
the ``exact'' calculation.  

Within the implementation of the ``point form'' approach used here, 
the only way to explain the above discrepancies relies on large 
contributions from two-body currents. These ones have been derived 
to ensure both current conservation and to reproduce the power law 
behavior of the Born amplitude. This has been done  consistently 
with the choice for the single-particle current (the simplest one)  
and for an interaction model which allows for a derivation 
in closed form (the interaction only appears at first order). 
We will not comment on their performance with respect to 
the above requested properties since they have been derived in such 
a way that they should be fulfilled. It is more interesting to investigate the 
qualitative and quantitative features that these two-body 
currents evidence.

What characterizes the two-body currents considered here is that 
they have not much to do with standard ones. Even though they involve 
the exchange of a neutral boson, and despite the fact that the 
interaction model has been chosen to make them as simple as possible, 
they are considerably more complicated than similar currents in a 
non-relativistic approach. Obtained somewhat by ``brute force'', 
these currents cannot be related to time-ordered diagrams. 
The reason for this is simple. The initial and final states being 
described on different hyper-planes, corresponding therefore to different
invariant times, there is no way to define time-ordered diagrams 
unambiguously. The situation is different for the two-body currents 
ensuring the right $F_0/F_1$ ratio. Treated very approximately, 
these currents should also be a part of other relativistic quantum 
mechanics approaches. 

To ensure current conservation, a recipe is sometimes used in the 
literature~\cite{Allen:2000ge} that involves a pole at $Q^2=0$. 
This can be acceptable in the case where the spectrum in the $t$-channel 
exhibits a zero-mass particle, as it would be the case for the axial current 
in hadronic physics, where this particle could be 
the pion in the chiral symmetry limit. The two-body 
currents derived here do not evidence such a pole, as expected. 
Actually, using the above recipe, one could construct a single- 
and a two-body current separately conserved. When the equation 
of motion  is used however, it turns out that the terms with 
a pole at $Q^2=0$ cancel each other. This is an important constraint 
on the derivation of the two-body currents. 

Quantitatively, for a rather moderately bound system ($\bar{p}^2/m^2=0.2$), 
we found that two-body currents motivated by current 
conservation produce contributions  to elastic charge form factors 
ranging from  0\%   at low $Q^2$ up to 100\% at $Q^2=100\,m^2$. 
The situation is slightly different for an inelastic 
transition where there is no constraint like charge conservation which 
imposes corrections to vanish at $Q^2=0$. There are other constraints 
and corrections that can reach 100\% with a destructive character so that to 
recover relations such as $F_1(Q^2) \rightarrow 0$ when $Q^2 \rightarrow 0$ 
or  $F_2(Q^2)/ F_1(Q^2) \rightarrow Q^{-2}$ at high $Q^2$. Otherwise 
corrections vary from 25\% at low $Q^2$ for $F_2(Q^2)$ up to 100\% 
at $Q^2=100\,m^2$ for $F_1(Q^2)$. Corrections due to two-body currents 
motivated by 
the Born amplitude are important too and are the dominant ones beyond 
$Q^2=10\,m^2$. If one puts apart aspects specific to the ``point form'' 
approach (like the faster drop off), large similarities with a 
Galilean-type calculation are  observed. Evidently, larger effects may 
be obtained in a stronger bound system with a larger value of $\bar{p}^2/m^2$, 
as evidenced by some results for the case $M=0.1\,m$. 

In comparison with an ``exact'' calculation, present ``point form'' results with 
incorporation of contributions due to two-body currents still show 
striking discrepancies. The increase of the charge radius in the limit 
$M  \rightarrow 0$ largely persists. While the form factor at high $Q^2$ 
has the right power law, the relative strength is found to be too small by 
a factor of the order $(M/(2\bar{e}))^4$. Both effects can be ascribed 
to the fact that the dependence on $Q^2$ appears only through the factor
$v^2=(Q^2/(4\,M^2+Q^2))$ and involve interaction effects  that make 
$M\neq(2\bar{e})$. This suggests that significant two-body currents are still
missing. Curiously, an instant-form calculation in the limit of a large 
momentum of the system evidences similar features~\cite{Amghar:2002jx}. 
To recover 
the ``exact'' results in this case, contributions from two-body currents were 
needed. These ones, however, had a non-trivial structure, with an integrand 
behaving like $0/0$ in the large momentum limit. It seems that similar 
two-body currents are needed in the present case too. Having a non-perturbative 
character (their form reminds one of zero-mode contributions to the pion 
form factor in the light-front approach), they could not be obtained 
from minimal conditions such as current conservation or reproducing 
the Born amplitude. Studying these currents and accounting for them may 
be a task for the future.

\begin{acknowledgments}
%\begin{acknowledgement}
\noindent
We are grateful to S. Noguera for his encouragements throughout this work. 
We would also like to thank R. Medrano for providing us with unpublished work.
This work was supported in part by the EC-IHP Network ESOP
under contract HPRN-CT-2000-00130, MCYT (Spain) under
contract BFM2001-3563-C02-01 and IN2P3 - CICYT (PTh02-1).
%\end{acknowledgement}
\end{acknowledgments}

%%%%%%%%%%%%%%%%%%%%%%%%%%%%%%%%%Appendice%%%%%%%%%%%%%%%%%%%%%%%%%%%%%%%%%%%%%%
%%%%%%%%%%%%%%%%%%%%%%%%%%%%%%%%%%%%%%%%%%%%%%%%%%%%%%%%%%%%%%%%%%%%%%%%%%%%%%%%
%%%%%%%%%%%%%%%%%%%%%%%%%%%%%%%%%%%%%%%%%%%%%%%%%%%%%%%%%%%%%%%%%%%%%%%%%

\appendix

\begin{widetext}

\section{Form factors in the Wick-Cutkosky model}
\label{app:a}

\noindent
This appendix contains expressions of form factors in the Wick-Cutkosky model 
which are used in the present work, successively for a scalar and 
a vector probe.

\subsection{Matrix element of a scalar current}

\noindent
$\bullet$ Matrix element of the current $n=1 \rightarrow n=1$:
\begin{eqnarray}
F_0(q^2)&=&\frac{i}{2}\int \frac{d^4p}{4\pi}\,dz_f\,dz_i
 \frac{  (p^2-m^2)\,g_1(z_f)\,g_1(z_i) }{
\Big(m^2 - (P_f^2-2P_f\ccdot p)\,\frac{1+z_f}{2} -p^2 -i\epsilon\Big)^3\,
\Big(m^2 - (P_i^2-2P_i\ccdot p)\,\frac{1+z_i}{2} -p^2 -i\epsilon\Big)^3  }
\nonumber \\
&=&\frac{3\pi}{8} \int dz_f\,dz_i\,dx 
\,\frac{g_1(z_f)\,g_1(z_i)\,x^2\,(1-x)^2}{D^4}
\, I_0, 
\label{appa1}
\end{eqnarray}
with
\begin{eqnarray}
I_0 &=&  \frac{1}{2} \left( m^2 -\frac{1}{4}\, \left(Z_f+Z_i\right) \,
   \left(M_f^2\,Z_f+M_i^2 \, Z_i\right) 
  -\frac{1}{4}\,Q^2\,Z_f\,Z_i\right)   + \frac{1}{3} D, 
\nonumber \\
D &=& m^2-\frac{1}{4}\,\left( 2-Z_f-Z_i \right) \, 
\left(M_f^2\,Z_f+M_i^2\,Z_i\right) +\frac{1}{4}\,Q^2\,Z_f\,Z_i, 
\label{appa2}
\end{eqnarray}
where $Z_i=(1+z_i)\,x$ and $Z_f=(1+z_f)\,(1-x)$.

\noindent
$\bullet$ Matrix element of the current $n=1, l=0 \rightarrow n=2, l=0$:
\begin{eqnarray}
F_0(q^2)&=&\frac{i}{2}\int \frac{d^4p}{4\pi}\,dz_f\,dz_i
\frac{ \Big((\frac{1}{2}P_f-p)^2+m^2-\frac{1}{4}P_f^2\Big) \, (p^2-m^2) 
\,g_2(z_f)\,g_1(z_i)  }{
\Big(m^2 - (P_f^2-2P_f\ccdot p)\,\frac{1+z_f}{2} -p^2 -i\epsilon\Big)^4\,
\Big(m^2 - (P_i^2-2P_i\ccdot p)\,\frac{1+z_i}{2} -p^2 -i\epsilon\Big)^3  }
\nonumber \\
&=&\frac{\pi}{4} \int dz_f\,dz_i\,dx 
\,\frac{g_2(z_f)\,g_1(z_i)\,(1-x)^3\,x^2}{D^5}
\,\, I_0, 
\label{appa3}
\end{eqnarray}
with
\begin{eqnarray}
\label{appa4}
\nonumber
&I_0= \frac{1}{2}( 2\, A\,B + D\,(C-A-B) +D^2) , & 
H=\frac{1}{4}\left(M_f^2(Z_f+Z_i/2)+M_i^2\,Z_i/2\right)
+\frac{1}{8}\,Q^2\,Z_i\\
&A=\frac{1}{4}\left(Z_f+Z_i\right)\,\left(M_f^2\,Z_f+M_i^2\,Z_i\right) 
+\frac{1}{4}\,Q^2\,Z_f\,Z_i -m^2,&\quad B=A - 2\,H + 2\,m^2,
\quad C=-A + H - m^2. 
\end{eqnarray}
The quantity $D$, referred to here and below, is defined in eq. (\ref{appa2}).

%%%%%%%%%%%%%%%%%%%%%%%%%bbbbb

\subsection{Matrix element of a vector current}

\noindent
$\bullet$ Matrix element of the current $n=1 \rightarrow n=1$:
\begin{eqnarray}
\label{appb1}
\nonumber 
\lefteqn{
F_1(q^2)\,(P^{\mu}_f+P^{\mu}_i) + F_2(q^2)\,q^{\mu} = } \\
&&i\int \frac{d^4p}{4\pi}\,dz_f\,dz_i  
\frac{ \Big(  P_f^{\mu}+P_i^{\mu}-2\,p^{\mu}\Big) \, 
(p^2-m^2) \,g_1(z_f)\,g_1(z_i)}{
\Big(m^2 - (P_f^2-2P_f\ccdot p)\,\frac{1+z_f}{2} -p^2 -i\epsilon\Big)^3\,
\Big(m^2 - (P_i^2-2P_i\ccdot p)\,\frac{1+z_i}{2} -p^2 -i\epsilon\Big)^3  }
\nonumber \\
&&\qquad\qquad\qquad=\frac{3\pi}{8} \int dz_f\,dz_i\,dx 
\,\frac{g_1(z_f)\,g_1(z_i)\,(1-x)^2\,x^2}{D^4}
\, \Big(I_1\,(P_f^{\mu}+P_i^{\mu})\,+\, I_2\,q^{\mu}\Big),
\end{eqnarray}
with $q^{\mu}=(P_f-P_i)^{\mu}$ and
\begin{equation}
\label{appb2}
I_1=   \left( 2-Z_f-Z_i\right) \,
  \left(m^2-\frac{1}{4}\,M_f^2\,Z_f-\frac{1}{4}\,M_i^2 \, Z_i\right)    
-\frac{1}{3}\,D,  \quad I_2=-\left( Z_f-Z_i \right) 
\left(m^2-\frac{1}{4}\,M_f^2Z_f-\frac{1}{4}\,M_i^2Z_i\right). 
\end{equation}

\noindent
$\bullet$ Matrix element of the current $n=1, l=0 \rightarrow n=2, l=0$:
\begin{eqnarray}
\label{appb3}
\nonumber
\lefteqn{
F_1(q^2)\,(P^{\mu}_f+P^{\mu}_i) + F_2(q^2)\,q^{\mu}=} \\
&&i\int \frac{d^4p}{4\pi}\,dz_f\,dz_i
\frac{ \Big(  P_f^{\mu}+P_i^{\mu}-2\,p^{\mu}\Big) \,
\Big((\frac{1}{2}P_f-p)^2+m^2-\frac{1}{4}P_f^2\Big) \,
(p^2-m^2) \,g_2(z_f)\,g_1(z_i) }{
\Big(m^2 - (P_f^2-2P_f\ccdot p)\,\frac{1+z_f}{2} -p^2 -i\epsilon\Big)^4\,
\Big(m^2 - (P_i^2-2P_i\ccdot p)\,\frac{1+z_i}{2} -p^2 -i\epsilon\Big)^3  }
\nonumber \\
&&\qquad\qquad\qquad=\frac{\pi}{4} \int dz_f\,dz_i\,dx 
\,\frac{g_2(z_f)\,g_1(z_i)\,(1-x)^3\,x^2}{D^5}
\, \Big(I_1\,(P_f^{\mu}+P_i^{\mu})\,+\, I_2\,q^{\mu}\Big), 
\end{eqnarray}
with
\begin{eqnarray}
\nonumber 
I_1=P_C +\frac{1}{2}\, P_A -\frac{1}{2}(Z_f+Z_i)\, (P_C+P_A+P_B), \qquad
I_2=\frac{1}{2}\, P_A -\frac{1}{2}(Z_f-Z_i)\,(P_C+P_A+P_B),\\ 
P_A=\frac{1}{2}\,D\,(D-A),
\qquad P_B=\frac{1}{2}\,D\,(D-B), \qquad P_C=2\,A\,B  +  D\,(C-A-B) + D^2.
\label{appb4}
\end{eqnarray}
where $A$, $B$ and $C$ are defined in eq.~(\ref{appa4}).

%%%%%%%%%%%%%%%%%%%%%%%%%%%%%%cccccc
\section{Analytic expressions for form factors in the ``point form'' approach}
\label{app:c}

\noindent
Expressions of form factors calculated in the ``point form'' approach
for a simple interaction model v0, first 
presented in ref.~\cite{Desplanques:2001zw}, 
are given together with  details concerning their derivation 
as well as the non-relativistic expressions.

For the wave functions  we use solutions obtained with a Coulomb potential,
\begin{equation}
\label{appc0}
\left[ 4\left( p^2 + m^2\right) - M^2 \right] \Psi(\vec{p}\,) = 
- 4m \int  \frac{d \vec{p}\,'}{(2\pi)^3} \, 
V_{int}(\vec{p},\vec{p}\,')\,  \psi(p'), \qquad {\rm with} \qquad
V_{int}(\vec{p},\vec{p}\,') = -\frac{g^2}{(\vec{p}-\vec{p}\,')^2}.
\end{equation}
The wave functions of the ground and first excited states, 
$\phi_i(\vec{p}\,)$ and $\phi_f(\vec{p}\,)$, respectively, are then given by:
\begin{equation}
\label{appc1}
\phi_i(\vec{p}\,)=\sqrt{4 \, \pi} \,\frac{4 \, 
\kappa^{5/2}}{(\kappa^2+\vec{p}^{\,2})^2} ,\qquad\qquad
\phi_f(\vec{p}\,)= \sqrt{4\pi} \,
\frac{8 \kappa^{*5/2}}{(\kappa^{*2}+\vec{p}^{\,2})^3}\,
(\vec{p}^{\,2}-\kappa^{*2}),
\end{equation}
where $\kappa^2=m^2-\frac{1}{4}M^2$, $\kappa^{*2}=m^2-\frac{1}{4}M^{*2}$, the 
total mass $M\,(M^{*})$ being the one obtained from the Bethe-Salpeter equation 
($\kappa^2 \simeq 4\,\kappa^{*2}$). The binding energy $E$, referred to in
tables~\ref{t10}-\ref{t60}, is related to the total mass of a state by
$M^2=(2\,m-E)^2$. It has  been shown that an equation like eq.~(\ref{appc0})
reproduces rather well the (normal) spectrum of the Wick-Cutkosky model,
provided an effective coupling is used~\cite{Amghar:1999ii,Amghar:2000pc}.  
In particular, both models exhibit the same degeneracy pattern.
The wave functions of eqs.~(\ref{appc1}) should therefore be a good zeroth 
order approximation for our study, including for the extreme case $M^2=0$. 

With these wave functions, some of the form factors can be calculated 
analytically.  Thus, in the ``point form'' approach, the elastic form factor 
reads:
\begin{equation}
F_1(q^2=-Q^2)=  \frac{ 1+2 \frac{Q^2}{4M^2} }{
\left(1+ \frac{Q^2}{4M^2}\right)^4 
\left(1+\frac{Q^2}{16\kappa^2\left(1+ \frac{Q^2}{4M^2}\right)} \right)^2}, 
\qquad \qquad F_2(q^2=-Q^2)=0.
\label{appc2}
\end{equation}
Interestingly, the quantity $\frac{Q^2}{16 \kappa^2}$ at the denominator of 
$F_1(q^2)$ is divided by the factor $1+ \frac{Q^2}{4M^2} $. This one was 
introduced in many calculations to account for the Lorentz-contraction 
effect~\cite{Friar:1973,Friar:1976kj} but it turned out to be valid only at 
small $Q^2$. 
The inelastic form factors for a transition from the ground- to the first 
excited radial state, $\tilde{F}_1(q^2)$ and $\tilde{F}_2(q^2)$ are given by:
\begin{eqnarray}
\tilde{F}_1(q^2)= \sqrt{2}\, \frac{64 \,\kappa^4 
\,v^2\,(16\,m^2-4\kappa^2(1-v^2))}{(9 
\kappa^2+ v^2\,(16\,m^2-10\kappa^2)  +v^4\kappa^2)^3}\,(1+v^2) \,(1-v^2)^3, 
\nonumber \\
\tilde{F}_2(q^2)=\sqrt{2}\, \frac{64\,(3+v^2) \, \kappa^6 }{(9 \kappa^2+  
v^2\,(16\,m^2-10\kappa^2)  +v^4\kappa^2  )^3}\,(1+v^2) \,(1-v^2)^4,
\label{appc3}
\end{eqnarray}
where $v^2$ is defined after eq.~(\ref{2r}). There is no known analytic
expression for the form factor, $F_0(q^2=-Q^2)$.

In the non-relativistic case, the elastic and inelastic form factors are 
respectively given by:
\begin{eqnarray}
F_0(q^2)= F_1(q^2)=\frac{\kappa^4}{( \kappa^2+Q^2/16)^2},
\qquad\qquad\qquad\qquad
F_2(q^2)= 0,  \nonumber \\
F_0(q^2)=F_1(q^2)= \sqrt{2}\, \frac{64 \,\kappa^4 \, Q^2}{(9 
\kappa^2+Q^2)^3},\qquad\qquad
F_2(q^2)=\sqrt{2}\, \frac{192\, \kappa^6 }{(9 \kappa^2+Q^2)^3}.
\label{appc4}
\end{eqnarray}
The equality $\kappa=2\,\kappa^{*}=\frac{m\,\alpha_{eff}}{2}$ has been assumed. 
Taking into account that 
$M_f^2-M_i^2= 3 \, \kappa^2$, one can verify that the current conservation 
condition, eq.~(\ref{2b}), is fulfilled. 

To get analytic expressions for the form factors in the ``point form'' using 
Coulombian-type wave functions, the following relations have been employed:
\begin{eqnarray}
\label{appc5}
\nonumber
\lefteqn{
\int d \vec{p} \left(\kappa^2+p_x^2+p_y^2+(\frac{p_z-v\,e_p}{\sqrt{1-v^2}})^2  
\right)^{-1}
\left(\kappa^{*2}+p_x^2+p_y^2+(\frac{p_z+v\,e_p}{\sqrt{1-v^2}})^2  
\right)^{-1}=} \\ && \nonumber \qquad\qquad
\pi^2 \frac{(1-v^2)^{3/2}}{v} \left[ \frac{M+M^{*}}{M\,M^{*}}  
\arctan\left(\frac{v\,(M+M^{*})}{2\,(\kappa+\kappa^{*})}\right) - 
\frac{M-M^{*}}{M\,M^{*}}  
\arctan\left(\frac{v\,(M-M^{*})}{2\,(\kappa+\kappa^{*})}\right)  \right],  \\
\nonumber
\lefteqn{
\int d \vec{p} \left(\kappa^2+p_x^2+p_y^2+(\frac{p_z-v\,e_p}{\sqrt{1-v^2}})^2  
\right)^{-1} 
\left(\kappa^{*2}+p_x^2+p_y^2+(\frac{p_z+v\,e_p}{\sqrt{1-v^2}})^2  \right)^{-1} 
\frac{p_z}{e_p}=} \\ && \qquad\qquad
\pi^2 \frac{(1-v^2)^{3/2}}{v^2} \left[ \frac{M-M^{*}}{M\,M^{*}}  
\arctan\left(\frac{v\,(M+M^{*})}{2\,(\kappa+\kappa^{*})}\right) - 
\frac{M+M^{*}}{M\,M^{*}}  
\arctan\left(\frac{v\,(M-M^{*})}{2\,(\kappa+\kappa^{*})}\right)  \right]. 
\end{eqnarray}
Complete expressions for the form factors are obtained by adding appropriately 
derivatives of the above ones with respect to the quantities $\kappa^2$ or 
$\kappa^{*2}$, taking into account that $M$ and $M^{*}$ depend on them.

For the ground state,  the elastic form factor in the ``point form'' can be directly 
obtained from the non-relativistic one by making a change of variable, starting 
from the above expression:
\begin{eqnarray}
\label{appc6}
\nonumber
\lefteqn{
\int d\vec{p} \left(\kappa^2+p_x^2+p_y^2+(\frac{p_z-v\,e_p}{\sqrt{1-v^2}})^2  
\right)^{-1}
\left(\kappa^2+p_x^2+p_y^2+(\frac{p_z+v\,e_p}{\sqrt{1-v^2}})^2  \right)^{-1}= } 
\\ && \qquad\qquad (1-v^2)^{3/2} \int dp_x\,dp_y\,d\tilde{p}_z\, 
\left( (\kappa^2+p_x^2+p_y^2+v^2\, \tilde{m}^2)^2 
+2 \tilde{p}_z^2(\kappa^2+p_x^2+p_y^2-v^2\, \tilde{m}^2) +\tilde{p}_z^4
\right)^{-1}.
\end{eqnarray}
The integrand is identical to the non-relativistic one, where $m^2$ 
has been replaced by $\tilde{m}^2=m^2-\kappa^2$ ($=M^2/4$).

%%%%%%%%%%%%%%%%%%%%%%%ddddddd

\section{Two-body currents and form factors in a Galilean  approach}
\label{app:d}

\noindent
Expressions for the two-body currents in a relativized, Galilean invariant 
approach (interaction model v1 in the text), together with their 
contributions to form factors are given in this subsection. 
Their presentation roughly follows the same structure as those 
for the ``point form'' case, eqs.~(\ref{5w},\ref{5x}). 
\begin{eqnarray}
F_0(q^2)&=& \frac{\sqrt{N_f\,N_i}}{4\,m} 
\left(\int \frac{d \vec{p}\,' \, d \vec{p}}{(2\pi)^3} 
\,\phi_f(\vec{p}\,') \, \frac{2\,m}{e_p+e_{p'}} \, \phi_i(\vec{p}\,) 
\,\, \delta( \frac{1}{2}\vec{q}+\vec{p}\,'-\vec{p} )
+ \int \frac{ d\vec{p}\,'\,d\vec{p}}{(2\pi)^6}
\,  \phi_f(\vec{p}\,')\,\phi_i(\vec{p}\,) \, 
\frac{m^2}{e_p e_{p'}}\, (K_{\Delta B})_0 \right), \nonumber \\
F_1(q^2)&=& \frac{\sqrt{N_f\,N_i}}{2\,\bar{M}}\, 
 \left[   \left( 1- (\frac{\bar{M}-2\,\bar{e}}{\bar{M}+2\,\bar{e}})^2 \right)
 \int \frac{d\vec{p}\,'\, d\vec{p}}{(2\pi)^3} 
\,\phi_f(\vec{p}\,') \,\phi_i(\vec{p}\,)
\,\,\delta( \frac{1}{2}\vec{q}+\vec{p}\,'-\vec{p} )
\right. \nonumber \\ &&\left.
+    \left( 1- (\frac{\bar{M}-2\,\bar{e}}{\bar{M}+2\,\bar{e}})^2 \right)
\int \frac{ d\vec{p}\,'\,d\vec{p}}{(2\pi)^6}
\,  \phi_f(\vec{p}\,')\,\phi_i(\vec{p}\,) \, 
\frac{m^2}{e_p e_{p'}}\, (K_{int})_1
+\int \frac{ d\vec{p}\,'\,d\vec{p}}{(2\pi)^6}
\,  \phi_f(\vec{p}\,')\,\phi_i(\vec{p}\,) \, 
\frac{m^2}{e_p e_{p'}}\, (K_{\Delta B})_1  \right], \nonumber \\
F_2(q^2) \, \vec{q}&=& -\sqrt{N_f\,N_i}\, 
\left[  \left( 1- (\frac{\bar{M}-2\,\bar{e}}{\bar{M}+2\,\bar{e}})^2 \right) 
\int \frac{d \vec{p}\,'\,d \vec{p}}{(2\pi)^3} 
\, \phi_f(\vec{p}\,') \, \frac{\vec{p}+\vec{p}\,'}{e_p+e_{p'}} \,
\phi_i(\vec{p}\,)
\,\, \delta( \frac{1}{2}\vec{q}+\vec{p}\,'-\vec{p} )
\right. \nonumber \\ && \hspace{-2em}\left.
+ \left( 1- (\frac{\bar{M}-2\,\bar{e}}{\bar{M}+2\,\bar{e}})^2 \right) 
\int \frac{ d\vec{p}\,'\,d\vec{p}}{(2\pi)^6}
\,  \phi_f(\vec{p}\,')\,\phi_i(\vec{p}\,) \, 
\frac{m^2}{e_p e_{p'}}\, (\vec{K}_{int})_2 
+ \int \frac{ d\vec{p}\,'\,d\vec{p}}{(2\pi)^6}
 \phi_f(\vec{p}\,')\,\phi_i(\vec{p}\,) \, 
\frac{m^2}{e_p e_{p'}}\, (\vec{K}_{\Delta B})_2  \right],
\label{appd1}
\end{eqnarray}
where $\bar{M}=(M_i+M_f)/2$, $\bar{e}=(\bar{e}_i+\bar{e}_f)/2$. The 
normalization factors $N_{i,f}$ and the quantities $\bar{e}_{i,f}$ 
are defined by:
\begin{equation}
\frac{1}{N_{i,f}}=\int \frac{d \vec{p} }{(2\pi)^3} \, \phi^2_{i,f}(\vec{p}\,) \, 
\frac{4\,e_p}{(M+2\,e_p)^2}
=\frac{4\,\bar{e}_{i,f}}{(M+2\,\bar{e}_{i,f})^2}\,
\int \frac{d \vec{p} }{(2\pi)^3} \, \phi^2_{i,f}(\vec{p}\,).
\label{appd2} 
\end{equation}
The expressions of the $K$ quantities, which account for two-body currents, are
given by:
\begin{eqnarray}
\label{appd3}
(K_{int})_1&=&  0, \nonumber \\ 
(\vec{K}_{int})_2&=&  
-\frac{g^2}{ \mu^2+(\frac{1}{2}\vec{q}+\vec{p}\,'-\vec{p} )^2}
\, \frac{1 }{  e_{p'+q/2} \, 
e_{p-q/2}}\,\left( e_p \frac{\vec{p}\,'+\vec{q}/4}{e_{p'+q/2}+e_{p'}}\, + e_{p'} 
\frac{\vec{p}-\vec{q}/4}{e_{p-q/2}+e_p} 
 \right) , 
\nonumber \\
(K_{\Delta B})_0&=&
\frac{g^2}{ \mu^2+(\frac{1}{2}\vec{q}+\vec{p}\,'-\vec{p} )^2}
\left( \frac{ 2\,m }{2\, e_{p'+q/2} \, (e_{p'+q/2}+e_{p})  } 
+ \frac{2\,m }{2\,  e_{p-q/2} \, (e_{p-q/2} + e_{p'})  } \right),
\nonumber \\
(K_{\Delta B})_1&=& 
\frac{g^2}{ \mu^2+(\frac{1}{2}\vec{q}+\vec{p}\,'-\vec{p} )^2}
\left(\frac{ e_{p'} \, (e_{p'+q/2}-e_{p'}) }{
2\,e_{p'+q/2} \,  (e_{p'+q/2}+e_{p}) \, (e_{p'+q/2}+e_{p'}) } 
+ \frac{e_p \, (e_{p-q/2}-e_p) }{  
2\,e_{p-q/2} \,(e_{p-q/2} + e_{p'})  \, (e_{p-q/2}+e_p)  } 
 \right),  \nonumber \\
(\vec{K}_{\Delta B})_2&=& 
\frac{g^2}{ \mu^2+(\frac{1}{2}\vec{q}+\vec{p}\,'-\vec{p})^2}
\left(\frac{ e_{p'} \, q_0\, (\vec{p}\,'+\vec{q}/4)  }{
2\,e_{p'+q/2} \, (e_{p'+q/2}+e_{p}) \, (e_{p'+q/2}+e_{p'})^2 } 
+  \frac{e_p \, q_0 \, (-\vec{p}+\vec{q}/4) }{  
2\,e_{p-q/2} \, (e_{p-q/2} + e_{p'})  \, (e_{p-q/2}+e_p)^2  } 
\right), \nonumber \\
\end{eqnarray}
where $q^0$ is defined as $M_f-M_i$, consistently with 
a Galilean-invariant calculation. While one could recover 
eqs.~(\ref{5h},\ref{5i}) in some limit, the above expressions 
evidence significant differences. Equations (\ref{5h},\ref{5i}) 
contain denominators with four energy terms instead of two here. 
On the other hand, they do not contain a squared term 
at the denominator like in  eq.~(\ref{appd3}). 
These differences illustrate the absence of a guide to derive 
two-body currents as soon as an effective interaction is used. 
Notice also that the  term ensuring current conservation, 
$\vec{K}_{int}$, does not contain any $1/q^2$ term.

%%%%%%%%%%%%%%%%%%%%%%%%%eeeeeee

\section{Two-body currents in the ``point form'' approach} 
\label{app:e}

\noindent
The difference between the current in the full Born approximation 
and the one accounted for by  solving eqs.~(\ref{5j},\ref{5k}) 
is given by:
\begin{eqnarray}
\label{appe1}
\nonumber 
\lefteqn{
J^{\mu}_{\Delta B}(q,p_{1i},p_{2i},p_{1f},p_{2f})=}\\ \nonumber 
&& -\sqrt{\frac{m}{\lambda_f \ccdot p_{1f}}\,\frac{m}{\lambda_f \ccdot p_{2f}}} \, 
\frac{g^2 \, \delta(\vec{q}+\vec{P_i}-\vec{P_f})}{\mu^2-(p_{2i}-p_{2f})^2}\, 
\sqrt{\frac{m}{\lambda_i \ccdot p_{1i}}\,\frac{m}{\lambda_i \ccdot p_{2i}}} \,\,
\frac{2 \lambda_i^{\mu}\, (\lambda_i \ccdot  p_{2i}) - p^{\mu}_{2f} + 
2  \lambda_f^{\mu}\, (\lambda_f  \ccdot   
p_{2f}) - p^{\mu}_{2f}}{4\,\lambda_i  \ccdot   p_{2i} 
\,(\lambda_i  \ccdot   p_{2i}-\lambda_i  \ccdot  p_{2f}) } \\ \nonumber &&
+\sqrt{\frac{m}{\lambda_f \ccdot p_{1f}}\,\frac{m}{\lambda_f \ccdot p_{2f}}} \,
\frac{g^2\,\delta(\vec{q}+\vec{P_i}-\vec{P_f})}{\mu^2-(p_{2i}-p_{2f})^2 
  + (\lambda_i  \ccdot   (p_{2i}-p_{2f}) )^2 } \,
  \sqrt{\frac{m}{\lambda_i \ccdot p_{1i}}\,\frac{m}{\lambda_i \ccdot p_{2i}}} \,\,
\frac{2 \lambda_i^{\mu} (\lambda^i  \ccdot p_{2f}) - p^{\mu}_{2f} + 
2  \lambda_f^{\mu} (\lambda_f \ccdot p_{2f}) - p^{\mu}_{2f}}
{4\,\lambda_i \ccdot p_{2f} \,(\lambda_i \ccdot p_{2i}-\lambda_i 
 \ccdot p_{2f}) } \\ && 
 \qquad \, + \, (i \leftrightarrow f ). 
\end{eqnarray}
This can be rewritten in a way which emphasizes the absence of the pole term 
$1/(\lambda_i  \ccdot   p_{2i}-\lambda_i  \ccdot  p_{2f})$:
\begin{eqnarray}
\label{appe2}
\nonumber
\lefteqn{
 J^{\mu}_{\Delta B}(q,p_{1i},p_{2i},p_{1f},p_{2f})= 
 \sqrt{\frac{m}{\lambda_f \ccdot p_{1f}}\,\frac{m}{\lambda_f \ccdot p_{2f}}} \, 
\frac{g^2}{2}\,\delta(\vec{q}+\vec{P_i}-\vec{P_f}) \, 
\sqrt{\frac{m}{\lambda_i \ccdot p_{1i}}\,\frac{m}{\lambda_i \ccdot p_{2i}}}} 
\\ \nonumber &&
%} \\ && \times \,
\times\,\Bigg[ \,
\left( \frac{\lambda_f^{\mu} (\lambda_f\ccdot p_{2f})-p_{2f}^{\mu} 
}{\lambda_i\ccdot p_{2i}}+ 
\lambda_i^{\mu}\right)
\left( \frac{\lambda_i  \ccdot   (p_{2f}-p_{2i}) }{H(0) \,H(\lambda_i)} + 
\frac{(\lambda_i+\lambda_f ) \ccdot   (p_{2i}-p_{2f})
}{H(\lambda_f) \,H(\lambda_i)}  \right)  
 \\ &&  \qquad\qquad  +
\left( \frac{\lambda_i^{\mu} (\lambda_i\ccdot p_{2i})-p_{2i}^{\mu} 
}{\lambda_f\ccdot p_{2f}}+ 
\lambda_f^{\mu}\right)
\left( \frac{\lambda_f  \ccdot   (p_{2i}- p_{2f}) }{H(0) 
\,H(\lambda_f)} + 
\frac{(\lambda_i + \lambda_f)  \ccdot   (p_{2f}-  p_{2i})
}{H(\lambda_f) \,H(\lambda_i)}  \right)  \, \Bigg].
\end{eqnarray}
For the scalar probe, one gets similarly:
\begin{eqnarray}
\label{appe3} 
\nonumber
\lefteqn{
 S_{\Delta B}(q, p_{1i},p_{2i},p_{1f},p_{2f})=
-\sqrt{\frac{m}{\lambda_f \ccdot p_{1f}}\,\frac{m}{\lambda_f \ccdot p_{2f}}}\,
\frac{g^2\,\delta(\vec{q}+\vec{P_i}-\vec{P_f})}{\mu^2-(p_{2i}-p_{2f})^2} 
\frac{2 \,m}{4\,\lambda_i \ccdot p_{2i} \,(\lambda_i \ccdot p_{2i}-\lambda_i 
 \ccdot   p_{2f}) } \, 
 \sqrt{\frac{m}{\lambda_i \ccdot p_{1i}}\,\frac{m}{\lambda_i \ccdot p_{2i}}} } \\  
 \nonumber &&
+\sqrt{\frac{m}{\lambda_f \ccdot p_{1f}}\,\frac{m}{\lambda_f \ccdot p_{2f}}} \,
\frac{g^2\,\delta(\vec{q}+\vec{P_i}-\vec{P_f})}{\mu^2-(p_{2i}-p_{2f})^2
+(\lambda_i \ccdot (p_{2i}-p_{2f}) )^2 } \, 
\frac{2\,m}{4\,\lambda_i \ccdot p_{2f} \, 
(\lambda_i  \ccdot   p_{2i}-\lambda_i  \ccdot   p_{2f}) } \,  
\sqrt{\frac{m}{\lambda_i \ccdot p_{1i}}\,\frac{m}{\lambda_i \ccdot p_{2i}}} + 
(i \leftrightarrow f )  \nonumber \\ &&
= \sqrt{\frac{m}{\lambda_f \ccdot p_{1f}}\,\frac{m}{\lambda_f \ccdot p_{2f}}} \, 
\frac{g^2}{2}\,\delta(\vec{q}+\vec{P_i}-\vec{P_f}) \, 
\sqrt{\frac{m}{\lambda_i \ccdot p_{1i}}\,\frac{m}{\lambda_i \ccdot p_{2i}}}
\nonumber \\  && \qquad \times  \,
\left[
\,\frac{1}{H(0)} \,
\left( \frac{m}{\lambda_i  \ccdot   p_{2i}\,\lambda_i  \ccdot   p_{2f}\,} 
+ \frac{m}{\lambda_f  \ccdot   p_{2f}\,\lambda_f  \ccdot   p_{2i}\,} 
\right)   
 - \left( \frac{m \, \lambda_i  \ccdot   (p_{2i}-p_{2f}) }{\lambda_i  \ccdot   
p_{2f}\, H(0) \,H(\lambda_i)}
 + \frac{m \, \lambda_f  \ccdot   (p_{2f}- p_{2i}) }{\lambda_f  \ccdot   
p_{2i}\, H(0) 
\,H(\lambda_f)} \right) \,  \right].
\end{eqnarray} 
Contribution of a two-body current to the matrix element of 
the scalar or vector current:
\begin{eqnarray}
\label{appe4}
\nonumber 
\lefteqn{
\sqrt{2E_f\,2E_i} \,
\langle f| \left( \begin{array}{c}
                   S_{\Delta B}\\ J^{\mu}_{\Delta B} 
                \end{array} \right)|i\rangle = } \\ &&
\sqrt{N_f\,N_i}  \frac{1}{(2\pi)^6 }  
 \int d^4p_{1f} \, d^4p_{1i}\, d^4p_{2f} \, d^4p_{2i} \, d\eta_f \, d\eta_i \, 
\delta(p_{1f}^2-m^2) \, \delta(p_{1i}^2-m^2)\,  \delta(p_{2f}^2-m^2) \, 
\delta(p_{2i}^2-m^2)  
\nonumber \\ &&
\times \,\theta( \lambda_f \ccdot   p_{1f}) \, 
\theta(\lambda_f \ccdot   p_{2f}) 
\,\theta(\lambda_i \ccdot   p_{1i})\, 
\theta(\lambda_i \ccdot   p_{2i}) \, 
\delta^4(p_{1f}+p_{2f}-\lambda_f \eta_f) 
\,\delta^4(p_{1i}+p_{2i}-\lambda_i \eta_i)  \nonumber \\ &&
\times \,\phi_f((\frac{p_{1f}-p_{2f}}{2})^2)\, 
\phi_i((\frac{p_{1i}-p_{2i}}{2})^2) \,
\sqrt{(p_{1f}+p_{2f})^2 \, (p_{1i}+p_{2i})^2 } \,\,4\,m^2
\left( \begin{array}{c}
                   \tilde{S}_{\Delta B}\\ \tilde{J}^{\mu}_{\Delta B}
                \end{array} \right)(q,p_{1f}, p_{2f},p_{1i},p_{2i}),
\end{eqnarray}
where $\tilde{S}_{\Delta B}\,(\tilde{J}^{\mu}_{\Delta B})$ represents 
the quantity $S_{\Delta B}\,(J^{\mu}_{\Delta B})$ of 
eqs.~(\ref{appe2},\ref{appe3}) excluding the $\delta$ function 
and the normalization factors $m/e$ accounted for separately.

Corresponding one-body contribution (differs from eq.~(\ref{2p}) 
by the presence of the overall factor $\sqrt{N_f\,N_i}$ in place of 
$\sqrt{2\,M_f\,2\,M_i}$):
\begin{eqnarray}
\label{appe5} 
\nonumber 
\sqrt{2E_f\,2E_i} \,\left<f|S|i\right> &=& \sqrt{N_f\,N_i} \, 
\frac{1}{(2\pi)^3 } 
  \int d^4p \, d^4p_f \,  d^4p_i \, d\eta_f \, d\eta_i \, 
\delta(p^2-m^2) \,  \delta(p_f^2-m^2) \, \delta(p_i^2-m^2)  \\ &&
\times \,\theta( \lambda_f \ccdot   p_f) \, 
\theta(\lambda_f \ccdot   p) 
\,\theta(\lambda_i \ccdot   p)\, 
\theta(\lambda_i \ccdot   p_i) \, 
\delta^4(p_f+p-\lambda_f \eta_f) 
\,\delta^4(p_i+p-\lambda_i \eta_i) \nonumber \\ && \qquad\qquad
\times \,\phi_f((\frac{p_f-p}{2})^2)\, \phi_i((\frac{p_i-p}{2})^2) \,
\sqrt{(p_f+p)^2 \, (p_i+p)^2 }\, 2m. 
\end{eqnarray}
\end{widetext}

%%%%%%%%%%%%%%%%%%%%%%%%%%%%%%%%%%%%%%%%%%%%%%%%%%%%%%%%%%%%%%%%%%%%%%%%%%%%%%
%%%%%%%%%%%%%%%%%%%%%%%%%%%%%%%%%%%%%%%%%%%%%%%%%%%%%%%%%%%%%%%%%%%%%%%%%%%%%%

%\bibliographystyle{elsart-num}
%\bibliographystyle{mybib}
%\bibliographystyle{myphysrev}
%\bibliographystyle{apsrev}
%\bibliography{2bcurrent} 

\end{document}